\documentclass[aps,prd,preprint,groupedaddress]{revtex4}

\usepackage{graphicx}
\usepackage{dcolumn}

\begin{document}

\preprint{UK/04-10}

\title{The Sequential Empirical Bayes Method: 
An Adaptive Constrained-Curve Fitting Algorithm for Lattice QCD}

\author{Ying Chen}
\author{Shao-Jing Dong}
\author{Terrence Draper}
\author{Ivan Horv\'{a}th}
\author{Keh-Fei Liu}
\author{Nilmani Mathur}
\author{Sonali Tamhankar}
\affiliation{Department of Physics and Astronomy, 
             University of Kentucky, 
             Lexington, KY 40506, USA}
\author{Cidambi Srinivasan}
\affiliation{Department of Statistics, 
             University of Kentucky, 
             Lexington, KY 40506, USA}
\author{Frank X. Lee}
\affiliation{Center for Nuclear Studies, 
             Dept.\ of Physics, 
             George Washington Univ.,
             Washington, DC 20052, USA}
\affiliation{Jefferson Lab, 
             12000 Jefferson Avenue, 
             Newport News, VA 23606, USA}
\author{Jianbo Zhang}
\affiliation{CSSM and Dept.\ of Physics and Math.\ Physics,
             Univ.\ of Adelaide, 
             Adelaide, SA 5005, Australia}


\begin{abstract}
We introduce the ``Sequential Empirical Bayes Method'', an adaptive
constrained-curve fitting procedure for extracting reliable priors.  These are
then used in standard augmented-$\chi^2$ fits on separate data.  This better
stabilizes fits to lattice QCD overlap-fermion data at very low quark mass
where {\it a priori\/} values are not otherwise known.  Lessons learned
(including caveats limiting the scope of the method) from studying artificial
data are presented.  As an illustration, from local-local two-point correlation
functions, we obtain masses and spectral weights for ground and first-excited
states of the pion, give preliminary fits for the $a_0$ where ghost states (a
quenched artifact) must be dealt with, and elaborate on the details of fits of
the Roper resonance and $S_{11}(N^{1/2-})$ previously presented elsewhere.  The
data are from overlap fermions on a quenched $16^3\times 28$ lattice with
spatial size $La=3.2\,{\rm fm}$ and pion mass as low as $\sim 180\,{\rm MeV}$.
\end{abstract}


\maketitle

\section{Introduction}

The recent advocacy of the use of Bayesian statistics for the analysis of data
from lattice simulations, in the guise of the methods of constrained curve
fitting~\cite{Lepage:2001ym,Morningstar:2001je}, or maximum
entropy~\cite{Asakawa:2000tr,Nakahara:1999vy,Fiebig:2002sp}, has eased
considerably the ambiguity and irritation associated with estimating the
systematic errors due to curve fitting, especially when extracting masses,
spectral weights and matrix elements from Monte Carlo estimates of correlation
functions.

Previously, Monte Carlo estimates, $\langle G(t) \rangle$, of two-point
hadronic correlators had been fit to a theoretical model, such as
\begin{eqnarray} \label{Exp}
      G(t;w_i,m_i) 
& = & 
      \sum_{i=1}^{\infty} w_i e^{-m_i t}
\end{eqnarray}
where $w_i$ is the spectral weight of the $i^{\rm th}$ state, by the
maximum-likelihood procedure of minimizing the $\chi^2$
\begin{eqnarray}\label{ChiSquare}
      \chi^2(w_{i},m_{i}) 
& = &
      \sum_{t,t^\prime}
        \left( \langle G(t)\rangle - G(t;w_i,m_i) \right)
        \sigma^{-2}_{t,t^\prime}
        \left( \langle G(t^\prime)\rangle - G(t^\prime;w_i,m_i) \right)
\end{eqnarray}
with covariance matrix      
\begin{eqnarray}\label{Covar}
      \sigma^2_{t,t^\prime} 
& = &   
      \langle G(t)G(t^\prime) \rangle - 
      \langle G(t) \rangle \langle G(t^\prime) \rangle
\end{eqnarray}
Traditionally, these had been fit only at large Euclidean times $t>t_{\rm
min}$, where contributions from excited states are exponentially damped.  The
art had been to choose a value of $t_{\rm min}$ which compromises between
unnecessarily high statistical errors for large $t_{\rm min}$ and high
systematic errors (from contamination from excited states) for small $t_{\rm
min}$.  Lattice alchemy provided various recipes for making the compromise and
estimating the systematic errors, but the procedures were often suspect and
always frustrating.

The truncation of the data set to only a few large $t$ was deemed necessary
because the alternative (of including more time slices but also more terms in
the fit model) resulted in unacceptably unstable fits to the sum of decaying
exponentials (traditionally a bane of numerical analysts).  Success was
achieved in some cases by enlarging the data set by including more channels,
e.g.\ diagonalization of multi-source multi-exponential fits.  Indeed when
correlators from very many sources could be calculated cheaply, such as for
glueballs or static quarks, the improvement was dramatic.  But most often, when
only a couple of channels at best could be fit simultaneously, the competition
between increased statistical errors for large $t_{\rm min}$ and large
systematic errors for small $t_{\rm min}$ remained; although the final
statistical and systematic errors were reduced, the effort and uncertainty in
obtaining a reliable systematic error remained.

Constrained curve fitting~\cite{Lepage:2001ym,Morningstar:2001je} offers the
alternative of minimizing an augmented $\chi^2$,
\begin{eqnarray}\label{AugmentedChiSquare}
	\chi_{\rm aug}^2 
& = &
	\chi^2 + \chi_{\rm prior}^2 \\
	\chi_{\rm prior}^2 
& = &
	\sum_i \frac{(\rho_i-\tilde{\rho}_i)^2}
	            {\tilde{\sigma}_i^2}            \label{ChiPrior}
\end{eqnarray}
where $\rho_i$ denotes the collective parameters of the fit (e.g.\ $\rho_i =
\{w_i,m_i\}$ for a sum of exponentials), as a way of achieving stability by
``guiding the fit'' with the use of Bayesian priors, that is, values of the
parameters obtained from {\it a priori\/} estimates $\rho_i =
\tilde{\rho}_i\pm\tilde{\sigma}_i$.  With improved stability, the data sets can
be enlarged to include small $t$ and the theory can be enlarged by including
many more terms in the fit model until convergence is obtained.  The systematic
error associated with the choice of $t_{\rm min}$ is thereby largely absorbed
into the statistical error.

The advantage that the constrained curve fitting of lattice data has over a
typical data set that a numerical generalist would consider, is that often we
have reliable estimates of, or at least constraints on, the fit parameters from
outside the data (for example, the masses must be positive, or the level
spacing is expected to be such-and-such from reliable models) which can then be
used as Bayesian priors.  Examples, such as upsilon
spectroscopy~\cite{Lepage:2001ym} where the level spacing can be reliably
estimated from quark models and experiments, are impressive.  Remarkably,
constrained curve fitting with Bayesian priors on such data has been able to
give satisfactory fits for local-local correlation functions, i.e.\ when
multi-source fits are unavailable (presumably due to prohibitive cost).

But with our recent data~\cite{Dong:2003zf}, we enter previously unexplored
territory.  We work with overlap fermions with exact chiral symmetry at
unprecedented small quark mass and large spatial volume.  The literature, from
which to obtain estimates to be used as priors, is limited.  Furthermore, the
details of the level spacings (e.g.\ the Roper resonance and the
$\Lambda(1405)$) are hotly debated between advocates of quark models versus
those of chiral models.  The use of priors in standard constrained curve
fitting tends to ``lock in'' the fit (within a sigma or so); if one gets them
badly wrong, then the fitted results may be misleading.  Furthermore, the
stability of the fit results against choice of prior must be tested -- this
reintroduces an element of subjectivity.  As a modification of the basic
Bayesian-prior constrained-curve fitting (augmented $\chi^2$) procedure, we
propose to make it more automatic, and to further absorb systematic errors
associated with choice of prior into statistical errors.

In section II, we give an overview of the ``Sequential Empirical Bayes Method''
detailing our extension of constrained curve-fitting.  In Section III, we add
some further improvements to better assess and reduce systematic errors, and
study fits to artificial data where the true values of the parameters are
known.  In Section IV we give, as an illustration of the efficacy of the
algorithm, some results from our low quark mass overlap fermion data for the
excited states of the pion, present preliminary fits for the $a_0$ where ghost
states (a quenched artifact) must be dealt with, and comment on the details of
fits of the Roper resonance and $S_{11}(N^{1/2-})$ previously presented
elsewhere~\cite{Dong:2003zf}.  Our summary and conclusions follow in Section V.

\section{The Sequential Empirical Bayes Method}  \label{Sect:Sequential}

Bayesian statistics is an entire field in itself, with an old and broad
history; for an introduction, see~\cite{BayesianChoice,SubjectiveObjective}.
Empirical Bayes methods are intermediate between classical (``frequentist'')
and Bayesian methods.  The core ideas of a sequential analysis of the data
originate with Robbins~\cite{Robbins:1950,Robbins:1955,Robbins:1964}.  We
propose the Sequential Empirical Bayes (SEB) method as a refinement especially
well suited for the special properties of lattice Monte Carlo correlation
functions.

We begin in subsection~\ref{Sect:Synopsis} by giving a brief description of the
standard constrained-curve fitting approach emphasizing its basis in Bayesian
probability theory.  Our sketch relies heavily on a few recent and relevant
summaries~\cite{Lepage:2001ym,Morningstar:2001je,Fiebig:2002sp,NumericalRecipes};
indeed, our notation is an amalgam of theirs.  We follow with a descriptive
overview of our method in subsection~\ref{Sect:Overview} and a more detailed
rendering of the basic algorithm in subsection~\ref{Sect:BasicAlgorithm}.

\subsection{Synopsis of Standard Constrained-Curve Fitting}  \label{Sect:Synopsis}

To account for the more general case of multi-source fits, as well as to allow
the time dependence to be non-exponential (periodic boundary conditions result
in $\cosh$ or $\sinh$, and quenched artifacts can give rise to more complicated
temporal dependence), we write
\begin{eqnarray}\label{GeneralChiSquare}
      \chi^{2}(\rho)
& = &
      \sum_{\alpha\beta} 
        (M_{\alpha}(\rho) - D_{\alpha})
        \sigma^{-2}_{\alpha\beta}
        (M_{\beta}(\rho) - D_{\beta})
\end{eqnarray}
where $\rho$ are the collective parameters of the fit (e.g.\ $\rho_i =
\{w_i,m_i\}$ for a sum of exponentials), the indices $\alpha$, $\beta$
distinguish different values of the independent variable (time for correlation
functions) and different interpolating fields, $D$ are the Monte Carlo data,
$M_{\alpha}(\rho)$ is the fit model, and $\sigma^{2}_{\alpha\beta}$ is the
covariance matrix.

Minimizing the $\chi^2$ in Eq.~\ref{GeneralChiSquare} is the solution of the
problem of determining the set of fit parameters $\rho$ which maximizes
$P(D|\rho)$, the conditional probability of measuring the data $D$ given a set
of parameters, also known as the ``likelihood'' of the data.  Bayesian
inference turns this question around and demands that the solution of the
curve-fitting problem consist of determining the set of parameters $\rho$ which
maximizes $P(\rho|D)$, the conditional probability that $\rho$ is correct given
the measured data $D$.  That is, Bayesian inference asks which fit-model
parameters are most likely given the data.

The computation of the latter conditional probability is possible because of
the celebrated Bayes' theorem
\begin{eqnarray}\label{BayesTheorem}
            P(\rho|D) 
& = & 
            \frac{P(D|\rho) P(\rho)} 
		 {P(D)}             \\
& = & 
            \frac{P(D|\rho) P(\rho)} 
		 {\int d\rho\, P(D|\rho) P(\rho)}
\end{eqnarray}
which follows directly from the elementary properties of probability theory
\begin{eqnarray}\label{BayesProof}
      P(\rho|D) P(D) 
& = & 
      P(\rho\cap D) \\
& = & 
      P(D|\rho) P(\rho)
\end{eqnarray}
The unconditional probability $P(\rho)$ is the plausibility one assigns to the
parameters $\rho$ before the additional information of the fit is provided, and
is known as the ``Bayesian prior distribution''.  The conditional probability
$P(\rho|D)$ is known as the ``posterior probability distribution''; it is the
reassessment of the likelihood of the parameters after the fit incorporates the
data.  The unconditional probability $P(D)$, known as the ``prior predictive
probability'' of the data, is independent of $\rho$ and thus serves only as a
normalization constant; it is determined from Eq.~\ref{BayesTheorem} and the
normalization condition $\int d\rho P(\rho|D) = 1$.
 
Heuristically, Bayes' theorem can be thought of as ``posterior probability''
$\propto$ ``likelihood'' $\times$ ``prior probability''.  Traditional inference
(``frequentist theory'') uses the likelihood $P(D|\rho)$ to quantify all
statistical measures of inference (mean, standard deviation, $\chi^2$,
confidence limits, $\cdots$), while Bayesian inference uses the posterior
distribution $P(\rho|D)$ to determine the statistics.  From
Eq.~\ref{BayesTheorem}, Bayesian theory requires augmenting the information
that the frequentist theory uses with the additional estimate of the prior
distribution $P(\rho)$.  Of course, the two methods agree when the prior
distribution is chosen to be constant.

The likelihood $P(D|\rho)$, needed by both frequentists and Bayesians, can be
estimated if the Monte Carlo data set is sufficiently large.  Then, regardless
of the statistics of the underlying population distribution (the ensemble of
all possible configurations), the sample distribution (Monte-Carlo-generated
averages over finite sets of configurations) will have Gaussian statistics, as
assured by the Central Limit Theorem.  Thus
\begin{eqnarray}\label{CentralLimit}
      P(D|\rho)
& \propto &
      \exp(-\chi^2/2)
\end{eqnarray}
with the $\chi^2$ given by Eq.~\ref{GeneralChiSquare}.

The Bayesian prior distribution $P(\rho)$ is the probability that a particular
set of parameters $\rho$ is correct, {\it a priori\/} to the data analysis.  It
is implicit in that the Bayesian probabilities are conditional on some
background information.  Operationally, $P(\rho)$ serves as a constraint on the
parameters $\rho$.  For the physicist, these might be constraints on the ranges
of parametric values that are physically feasible (e.g ``positive energy
splittings''), or perhaps something stronger as dictated by experience from
previous similar fits or guidance from models of QCD or experiments.

A simple choice for the prior distribution is Gaussian
\begin{eqnarray}\label{GaussianPrior}
      P(\rho)
& \propto &
      \exp(-\chi_{\rm prior}^2/2)
\end{eqnarray}
with $\chi_{\rm prior}$ defined in Eq.~\ref{ChiPrior}.  Then $P(\rho|D)
\propto P(D|\rho) P(\rho)$ of Eq.~\ref{BayesTheorem} is maximized by
minimizing the augmented $\chi^2$ ($\chi_{\rm aug}^2 = \chi^2 + \chi_{\rm
prior}^2$) of Eq.~\ref{AugmentedChiSquare}.  This is the approach outlined
in~\cite{Lepage:2001ym,Morningstar:2001je} where they choose values for the
priors $\langle \rho \rangle = \tilde{\rho}$ and $\langle \rho^2 \rangle -
\langle \rho \rangle ^2 = \tilde{\sigma}^2$ based on physicists' intuition.

The opposite extreme is the use of the ``entropic prior'', based on the view
that for some fits where the number of parameters is larger than the number of
measured data (such as for the spectral density
function~\cite{Asakawa:2000tr,Nakahara:1999vy,Fiebig:2002sp}, or in a wider
context image restoration) only minimal information about the priors is
available.

We seek an approach between the two extremes.  We use the language of augmented
$\chi^2$ but seek to use a subset of the available data to estimate the priors
in an orderly fashion, which will then be used on new data.

\subsection{Overview of the SEB Method}  \label{Sect:Overview}

So how do we obtain our priors?  For concreteness, consider again the fit model
of the sum of decaying exponentials in Eq.~\ref{Exp}.  Apportion the data into
a nested set (picture the shells in an onion or a Kewpie doll) and extract
estimates for priors in a progressive manner, at each stage obtaining estimates
of one or two new priors upon each expansion of the data set.  An especially
natural and well-suited nesting is to partition the hadronic correlator data
via Euclidean time slices.  That is, the first and smallest data set includes
all configurations but with times slices $t$ restricted to $t\ge t_{\rm start}$
(and perhaps $t\le t_{\rm max}$ depending on boundary conditions).
Subsequently, the $n^{\rm th}$ data set includes all time slices such that
$t\ge t_{\rm start}-(n-1)\Delta t$, where $n=1,2,\cdots$ and $\Delta t$ is most
simply chosen to be the constant $2$. (A more general choice is made in
practice, as described later.)  The time $t_{\rm start}$ is chosen by the same
criteria used in traditional fits so that an unconstrained fit is well
approximated by a single exponential.  The output values for two parameters,
the ground state mass and spectral weight, are then used as priors for the next
fit on the augmented data set.  For the second fit, two (or more, in general)
additional time slices are included, as are two more parameters in the fit
model, the weight and mass of the first excited state.  The new fit is
constrained with regard to the ground state but unconstrained with regard to
the first excited state.  The four fitted parameters are then used as priors in
a third fit on a larger data set, and so on until all desired time slices
and/or terms in the fit are included.

In this way, we estimate the priors from a subset of the data.  Furthermore, we
choose that subset of the data which is best suited for making the estimation.
The choice is determined with the same compromise as is used for traditional
unconstrained fits, namely between minimizing statistical errors (which grow as
$t$ increases) and minimizing the contamination of higher excited states (which
grows as $t$ decreases).

Thus we propose an adaptive self-contained constrained curve-fitting procedure,
dubbed the ``Sequential Empirical Bayes'' method.  In a nutshell, we obtain the
priors gradually (allowing them to change as needed), from the ground state up,
as the data set is monotonically enlarged by including earlier and earlier time
slices.  Its advantages include that it is usable whenever external reliable
estimates of the priors are not available, and it is as automatic as one could
hope for, thereby reducing the potential to introduce bias and of course
decreasing the frustrating busy-work of fitting.  Especially for the low-lying
states, if the initial priors are estimated incorrectly, there are several
subsequent steps by which they may change (by about a sigma each time).

\subsection{The Basic Algorithm}\label{Sect:BasicAlgorithm}

From experience we have discovered that for some data, either ``real'' (actual
lattice gauge theory data) or artificially constructed, a very basic algorithm
which incorporates the SEB philosophy is adequate for producing reliable
priors.  However, some of our data have exposed deficiencies in the basic
algorithm.  These have been corrected without violating the spirit of the SEB,
but make the resulting final algorithm rather mysterious and complicated to
describe in one pass.  Accordingly, in this paper, we begin here by outlining
the simplest algorithm as a template.  In Sect.~\ref{Sect:Systematic} we will
introduce modifications as necessary.

\begin{table*}[ht]
\caption{\label{Table:Parameters} Table~\ref{Table:Parameters} The very basic
``fixed $\Delta t$'' algorithm, in which after each pair of steps, $\Delta t =
2$ new (earlier) time slices are added to the data to be fitted.  For
simplicity, the fit model of Eq.~\ref{Exp} is assumed in this example.  Refer
to the text for the meaning of the superscripts and subscripts on the masses
$m$ and spectral weights $w$.}
  \begin{tabular}{c|c|c|c|c}
\hline\hline
       & Time                          & Scanned                & Priors                             & Fitted               \\
Step   &      Slices                   &         Initial        &        (\& Other                   &        Output        \\
       &             Fitted            &                 Values &         Initial Values)            &               Values \\
\hline\hline
1      & \{$t_{\rm start}  ,t_{max}$\} & $w_{1}, m_{1}$         & --                                 & $w_{1}^{(1)}, m_{1}^{(1)}$, $\sigma_{w_{1}}^{(1)}$, $\sigma_{m_{1}}^{(1)}$ \\
2      & \{$t_{\rm start}-1,t_{max}$\} & --                     & $\tilde{w}_{1}^{(2)}=w_{1}^{(1)}$, 
                                                                  $\tilde{m}_{1}^{(2)}=m_{1}^{(1)}$  & $w_{1}^{(2)}, m_{1}^{(2)}$, $\sigma_{w_{1}}^{(2)}$, $\sigma_{m_{1}}^{(2)}$\\

\hline
3      & \{$t_{\rm start}-2,t_{max}$\} &                        & $\tilde{w}_{1}^{(3)}=w_{1}^{(2)}$, 
                                                                  $\tilde{m}_{1}^{(3)}=m_{1}^{(2)}$  & $w_{1}^{(3)}, m_{1}^{(3)}$, $\sigma_{w_{1}}^{(3)}$, $\sigma_{m_{1}}^{(3)}$\\
       &                               & $w_{2}, m_{2}$         &                                    & $w_{2}^{(3)}, m_{2}^{(3)}$, $\sigma_{w_{2}}^{(3)}$, $\sigma_{m_{2}}^{(3)}$\\
4      & \{$t_{\rm start}-3,t_{max}$\} & --                     & $\tilde{w}_{1}^{(4)}=w_{1}^{(3)}$, 
                                                                  $\tilde{m}_{1}^{(4)}=m_{1}^{(3)}$  & $w_{1}^{(4)}, m_{1}^{(4)}$, $\sigma_{w_{1}}^{(4)}$, $\sigma_{m_{1}}^{(4)}$\\
       &                               &                        & $\tilde{w}_{2}^{(4)}=w_{2}^{(3)}$, 
                                                                  $\tilde{m}_{2}^{(4)}=m_{2}^{(3)}$  & $w_{2}^{(4)}, m_{2}^{(4)}$, $\sigma_{w_{2}}^{(4)}$, $\sigma_{m_{2}}^{(4)}$\\

\hline
5      & \{$t_{\rm start}-4,t_{max}$\} &                        & $\tilde{w}_{1}^{(5)}=w_{1}^{(4)}$, 
                                                                  $\tilde{m}_{1}^{(5)}=m_{1}^{(4)}$  & $w_{1}^{(5)}, m_{1}^{(5)}$, $\sigma_{w_{1}}^{(5)}$, $\sigma_{m_{1}}^{(5)}$\\
       &                               &                        & $\tilde{w}_{2}^{(5)}=w_{2}^{(4)}$, 
                                                                  $\tilde{m}_{2}^{(5)}=m_{2}^{(4)}$  & $w_{2}^{(5)}, m_{2}^{(5)}$, $\sigma_{w_{2}}^{(5)}$, $\sigma_{m_{2}}^{(5)}$\\
       &                               &  $w_{3}, m_{3}$        &                                    & $w_{3}^{(5)}, m_{3}^{(5)}$, $\sigma_{w_{3}}^{(5)}$, $\sigma_{m_{3}}^{(5)}$\\

6      & \{$t_{\rm start}-5,t_{max}$\} & --                     & $\tilde{w}_{1}^{(6)}=w_{1}^{(5)}$, 
                                                                  $\tilde{m}_{1}^{(6)}=m_{1}^{(5)}$  & $w_{1}^{(6)}, m_{1}^{(6)}$, $\sigma_{w_{1}}^{(6)}$, $\sigma_{m_{1}}^{(6)}$\\
       &                               &                        & $\tilde{w}_{2}^{(6)}=w_{2}^{(5)}$, 
                                                                  $\tilde{m}_{2}^{(6)}=m_{2}^{(5)}$  & $w_{2}^{(6)}, m_{2}^{(6)}$, $\sigma_{w_{2}}^{(6)}$, $\sigma_{m_{2}}^{(6)}$\\
       &                               &                        & $\tilde{w}_{3}^{(6)}=w_{2}^{(5)}$, 
                                                                  $\tilde{m}_{3}^{(6)}=m_{2}^{(5)}$  & $w_{3}^{(6)}, m_{3}^{(6)}$, $\sigma_{w_{3}}^{(6)}$, $\sigma_{m_{3}}^{(6)}$\\
\hline
$\cdots$&                              &                        &                                    &                            \\
\hline
N      & \{$t_{\rm min},t_{max}$\}     & --                     & $\tilde{w}_{1}^{(N)}=w_{1}^{(N-1)}$, 
                                                                  $\tilde{m}_{1}^{(N)}=m_{1}^{(N-1)}$  & $w_{1}^{(N)}, m_{1}^{(N)}$, $\sigma_{w_{1}}^{(N)}$, $\sigma_{m_{1}}^{(N)}$\\
       &                               &                        & $\cdots$                             & $\cdots$                   \\
       &                               &                        & $\tilde{w}_{n}^{(N)}=w_{2}^{(N-1)}$, 
                                                                  $\tilde{m}_{n}^{(N)}=m_{2}^{(N-1)}$  & $w_{n}^{(N)}, m_{n}^{(N)}$, $\sigma_{w_{1}}^{(N)}$, $\sigma_{m_{1}}^{(N)}$\\
\hline\hline
  \end{tabular}
\end{table*}

Table~\ref{Table:Parameters} describes for simplicity the very basic ``fixed
$\Delta t$'' algorithm, in which after each pair of steps, $\Delta t = 2$ new
(earlier) time slices are added to the data to be fitted.  The algorithm is as
follows:

\begin{itemize}
\item Choose $t_{\rm max}$ and $t_{\rm min}$, the maximum window over which the
fits will be done.  The window is to be chosen as large as possible.  For
correlation functions expected, on theoretical grounds, to be positive then if
a value of a correlation function at a time slice is not within one sigma of
being positive, then that time slice and all greater time slices are eliminated
from consideration.  This prevents noisy correlation functions at large $t$
from giving grossly inaccurate estimates of the fitted values to be
subsequently used as priors.
\item Determine the number of the terms we want to use in the fit model, and
determine $t_{\rm start}$ as the starting point for the fit.  Ensure that
$(t_{\rm start}-t_{\rm min}) \geq \#\,{\rm terms} \times \Delta t$
\item Choose central trial values $w_{1}$ and $m_{1}$ (or $E_{1}$) equal to
those obtained from the effective mass relations.  Loop on various trial values
around these central values.  For each, use an unconstrained fit on the
one-mass-term model to fit the correlator data including time slices $t_{\rm
start}$ to $t_{\rm max}$ and obtain $w_{1}^{(1)}\pm \sigma_{w_{1}}^{(1)}$ and
$m_{1}^{(1)}\pm \sigma_{m_{1}}^{(1)}$.  Choose as input for the next step those
values which yield the lowest (but reasonable) $\chi^2/{\rm dof}$.
\item Using these values of $w_{1}\pm \sigma_{w_{1}}$ and $m_{1}\pm
\sigma_{m_{1}}$ as both priors and initial values, do a constrained curve fit
(using the one-mass-term model on the data set enlarged to include $t_{\rm
start} - 1$) to obtain $w_{1}^{(2)}\pm \sigma_{w_{1}}^{(2)}$ and
$m_{1}^{(2)}\pm\sigma_{m_{1}}^{(2)}$.
\item Loop on a wide range of trial values for $w_{2}$ and $m_{2}$.  With a
two-mass-term model, constrain the first mass and weight (using the previous
output as both priors and initial values) but leave the second mass and weight
unconstrained.  Loop on various trial values for the latter.  Do this
half-constrained fit on the data set enlarged to include $t_{\rm start} - 2$
and obtain $w_{2}^{(3)}\pm \sigma_{w_{2}}^{(3)}$ and
$m_{2}^{(3)}\pm\sigma_{m_{2}}^{(3)}$.  Choose as input for the next step those
values which yield the lowest (but reasonable) $\chi^{2}/{\rm dof}$.
\item Using these values of $w_{1}\pm\sigma_{w_{1}}$, $m_{1}\pm
\sigma_{m_{1}}$, $w_{2}\pm \sigma_{w_{2}}$, $m_{2}\pm\sigma_{m_{2}}$ as both
priors and initial values, do a fully-constrained fit (using the two-mass-term
model on the data set enlarged to include $t_{\rm start} - 3$) to obtain
$w_1^{(4)}\pm\sigma_{w_{1}}^{(4)}$, $m_{1}^{(4)}\pm \sigma_{m_{1}}^{(4)}$,
$w_{2}^{(4)}\pm \sigma_{w_{2}}^{(4)}$, and
$m_{2}^{(4)}\pm\sigma_{m_{2}}^{(4)}$.
\item Repeat the last two steps until all desired mass terms and time slices
are included.  One thus obtains a complete set of priors.
\item Add the final time slice $t_{\rm min}$ and do a fully-constrained fit
using previously obtained values for priors and initial guesses.
\end{itemize}

All fits are correlated using the full covariance matrix.  Furthermore, the
entire process is bootstrapped (or jackknifed).  Final quoted errors are
bootstrap (or jackknife) errors.  Within a bootstrap sample, at intermediate
steps in the algorithm, the sigmas of the priors, $\tilde{\sigma}_{\rho}$, are
obtained from the fitting errors of the previous step.

As a variation, rather than deciding {\it a priori\/} on the number of terms in
the fit and adding time slices one at time, one can let the data decide how
many time slices to include with each enlargement of the data by choosing the
minimum $\chi^{2}/{\rm dof}$ over a range of reasonable possibilities.  Thus,
for example, if the data is dominated by the ground state for many time slices,
then many time slices will be automatically added before an attempt is made to
fit the first-excited state.  We will refer to this improvement as the
``variable $\Delta t$'' approach, and to the basic template described in this
section as the ``fixed $\Delta t$'' approach.

\section{Assessing Systematic Errors} \label{Sect:Systematic}

The $\chi^{2}$ minimization can be cast more generally as minimizing a
functional ${\cal A}(\rho)>0$ of the vector $\rho$ (see, for
example,~\cite{NumericalRecipes}).  In the extreme case, with more unknown
parameters than data points, the minimization procedure becomes degenerate and
there is enough freedom to drive $\chi^{2}$ down to unreasonably small values.
With more data points than unknown parameters, although the agreement with the
data typically is good (often too good to believe), the solution is unstable
and often wildly oscillating.  This signals a proximity to the degeneracy of
the underlying minimization problem.  Constrained curve fitting can be cast
more generally as the problem of minimizing ${\cal A}(\rho) + \lambda {\cal
B}({\rho})$
\begin{eqnarray}\label{Minimization}
      \frac{\delta}{\delta {\bf \rho}}
      \left( {\cal A}(\rho) + \lambda {\cal B}({\rho}) \right)
& = & 
      0
\end{eqnarray}
If ${\cal A}(\rho)$ is degenerate, but ${\cal B}({\rho})$ is not, the
degeneracy is lifted.  The stability is improved in general.  The solution
$\rho({\lambda})$ varies along a ``trade-off curve'' (see, for
example,~\cite{NumericalRecipes}).  ${\cal A}$ (e.g.\ $\chi^2$) measures the
agreement of the model to the data, while ${\cal B}$ (e.g.\ $\chi_{\rm
prior}^2$) plays the role of a stabilizing functional.  The standard
constrained-curve fitting procedure selects $\lambda=1$.

To estimate the systematic errors associated with the choice of priors, it is
common to repeat the procedure with all $\tilde{\sigma}_{\rho}$ replaced by
$\eta \tilde{\sigma}_{\rho}$ where $\eta=2$ is a typical choice.  This moves
one along the trade-off curve from $\lambda=1$ to $\lambda=1/\eta^2$, giving
less weight to the priors.  More generally, one could have an independent
weighting factor, $\eta_{i}$, multiplying the $\tilde{\sigma}_{i}$ of each
prior.

\subsection{Bells and Whistles}

\subsubsection{``Global Dynamical Weight''}

For concreteness, we return to the simple case of a sum of decaying
exponentials of Eqs.~\ref{Exp},~\ref{ChiSquare}
\begin{eqnarray}\label{ChiSquarePriorEta}
	\chi_{\rm prior}^2 
& = &
	\sum_i \frac{(\rho_i-\tilde{\rho}_i)^2}
                    {\eta_i^2 \tilde{\sigma}_i^2}
\end{eqnarray}
By choosing $\eta_i^{2} > 1$, one relaxes the constraint of the priors and thus
tests the sensitivity of the fitted values to the choice of prior.  We present
three ways, in order of increasing sophistication, to implement the choice of
each $\eta_i$.

\begin{enumerate}

\item[(a)] {\it Global Static Weight:\/} The canonical approach is to keep it
unchanged as a global constant for the fit, $\eta_{i}^2=1/\lambda$.  The
variation of the dependence of the output fitted results on this global input
parameter may be used to estimate the systematic error.

\item[(b)] {\it Local Variable Weight:\/} Allow each $\eta_i$ to be a local
variable.  At each step, loop on various values of $\eta_i$ and choose the
value which minimizes the $\chi^{2}/{\rm dof}$.  Monitor against runaway
solutions.

\begin{figure}[ht]
\includegraphics[angle=0,width=0.45\hsize]{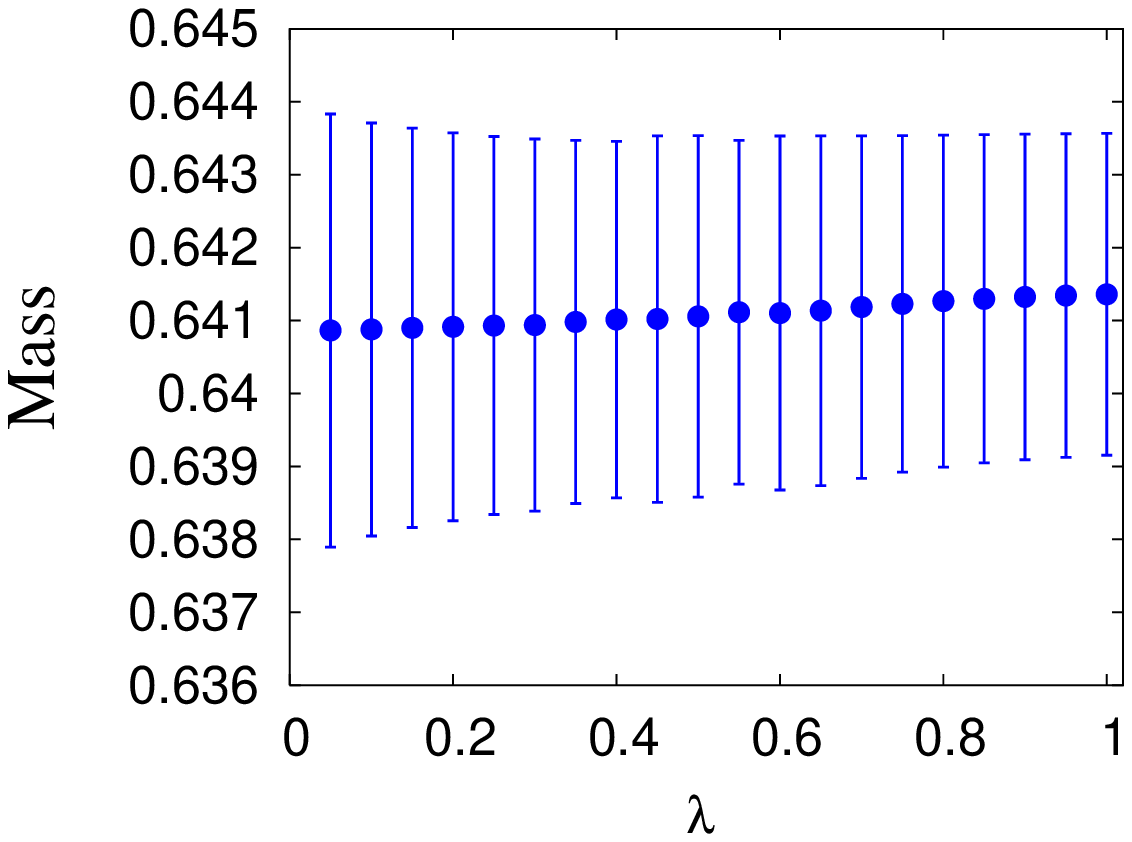}%
\includegraphics[angle=0,width=0.45\hsize]{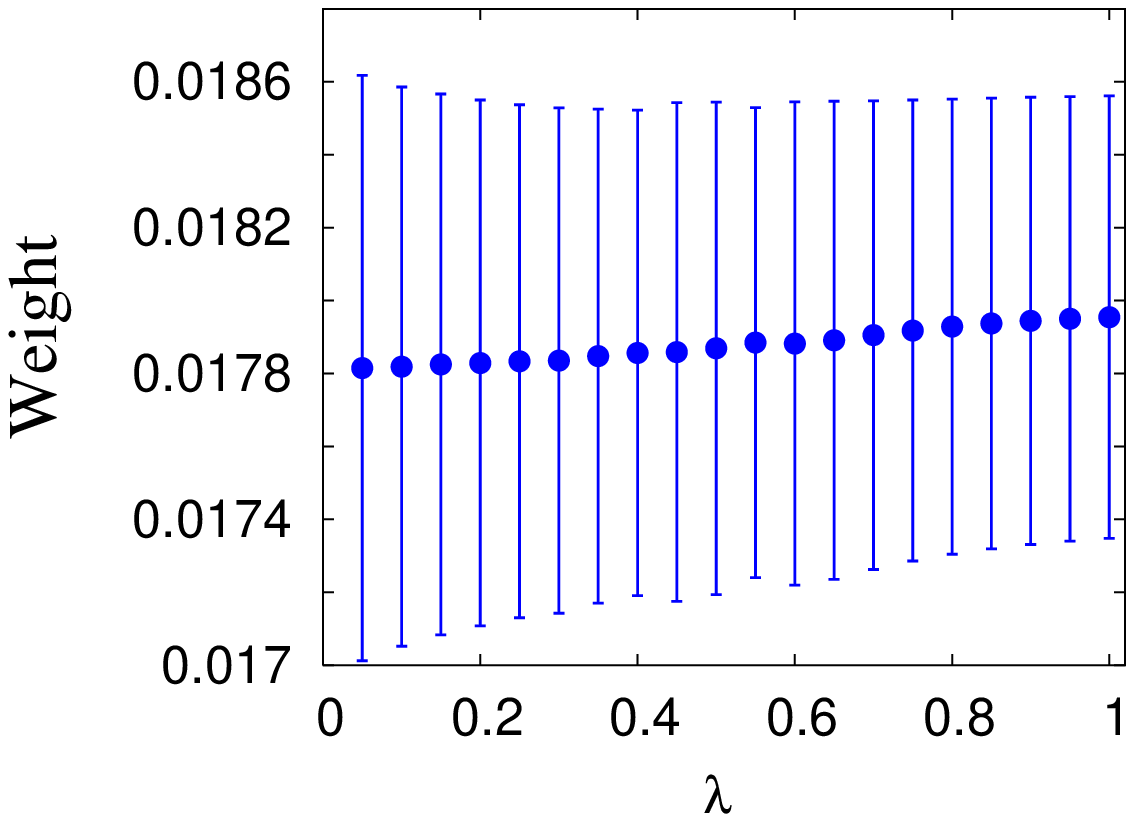}%
\caption{\label{Fig:PionLambdaTest} Plot of ground-state pion mass, $m_{\pi}a$
(left) and spectral weight $w$ (right), versus global static weight, $\lambda$,
for bare quark mass $ma=0.226$.}
\end{figure}

\begin{figure}[ht]
\includegraphics[angle=0,width=0.45\hsize]{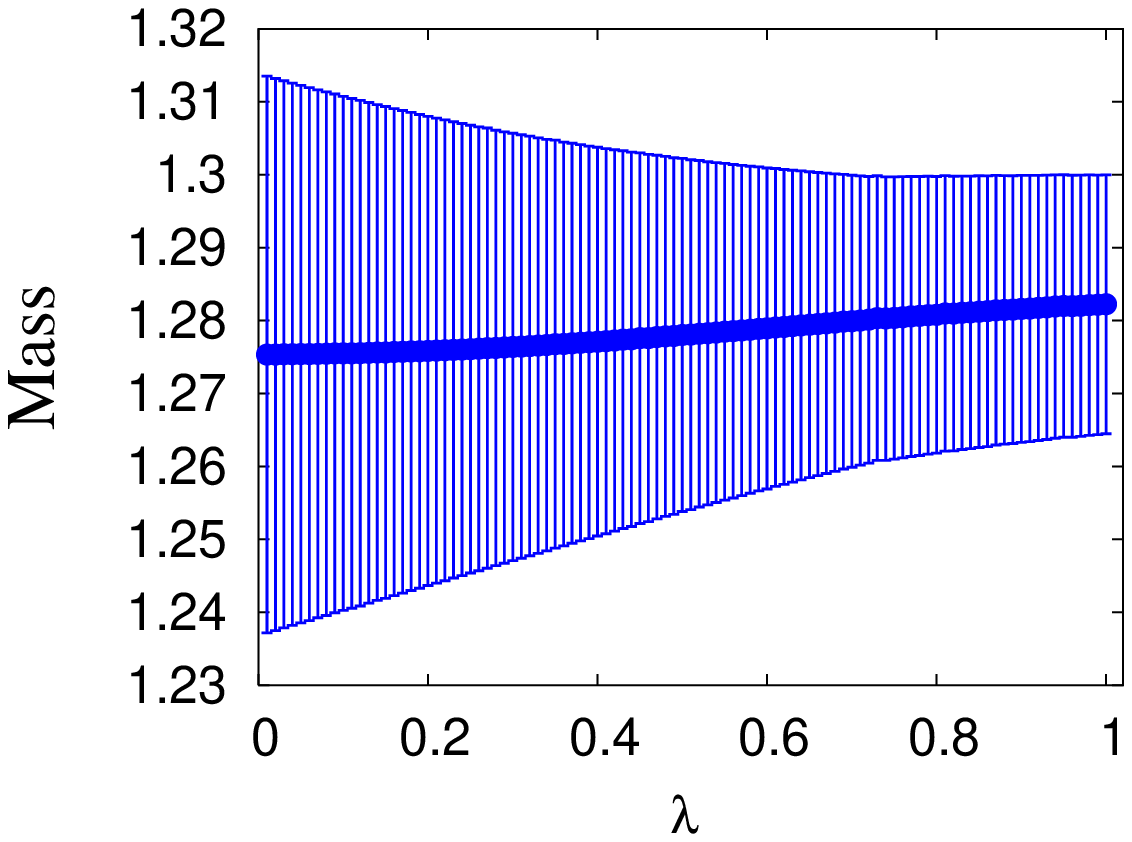}%
\includegraphics[angle=0,width=0.45\hsize]{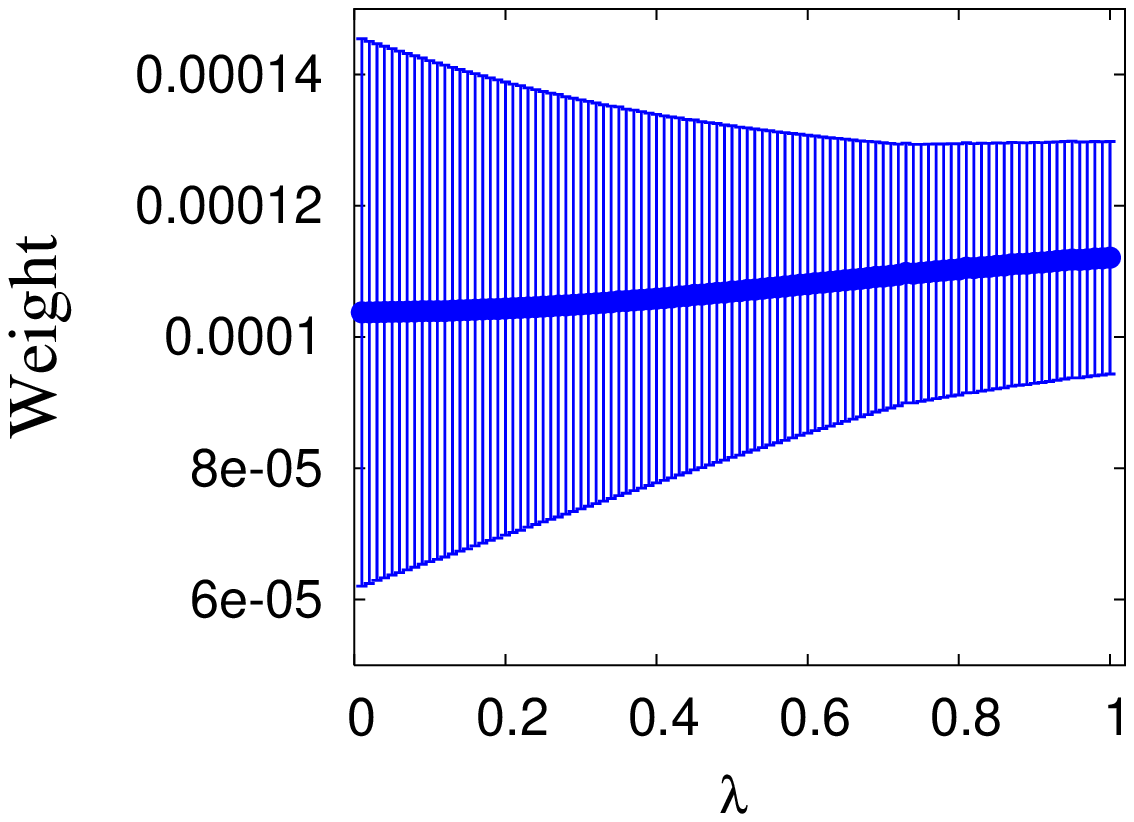}%
\caption{\label{Fig:NucleonLambdaTest} Same as in Fig.~\ref{Fig:PionLambdaTest}
but for the nucleon for bare quark mass $ma=0.188$ ($m_{\pi}=583(3)\,{\rm
MeV}$).}
\end{figure}

\item[(c)] {\it Local Dynamical Weight:\/} For some quantities, such as for the
ground-state mass and spectral weight of the pion displayed in
Fig.~\ref{Fig:PionLambdaTest}, the dependence of the fitted results on the
value of $\lambda$ is quite mild.  For others, such as for the nucleon of
Fig.~\ref{Fig:NucleonLambdaTest}, the dependence is stronger.  Accordingly, we
propose an adaptive procedure of absorbing any systematic error associated with
the choice of $\eta$ into the statistical error, by upgrading each $\eta_{i}$
to a dynamical variable to be determined by the fit.  It is incorporated as a
dynamical fit variable by further augmenting the $\chi^{2}$:
\begin{eqnarray}\label{ChiSquareAugEta}
	\chi_{\rm aug^{\prime}}^2 
& = & 
	\chi^2 + \chi_{\rm prior}^2 +\chi_{\eta}^2 \\
	\chi_{\rm prior}^2 
& = &	
	\sum_i \frac{(\rho_-\tilde{\rho}_i)^2}
	            {\eta_{i}^2\tilde{\sigma}_i^2} \\
	\chi_{\eta}^2 
& = & 
	\sum_i \frac{(\eta_{i} - \tilde{\eta_{i}})^2}
                    {\tilde{\sigma}_{\eta}^2}
\end{eqnarray}
The choice of $\tilde{\eta_i}$ and $\tilde{\sigma}_{\eta}$ is also somewhat
arbitrary, and results in some uncertainty in assessing the systematic error;
however, the dependence on these parameters is gentler than on the
corresponding static global values and any remaining systematic error is masked
by statistical error.

\item[(d)] {\it Global Dynamical Weight:\/} In our fits presented here, we made
the simplest choice of a special case of (c): we took all $\eta_i$ for
different parameters to be the same global $\eta$, and took $\tilde{\eta}=1$
and $\tilde{\sigma}_{\eta}=1$.

Often, the dependence of the results of the fit on the Global Static Weight is
sufficiently smooth that the use of Global Dynamical Weight is unnecessary.

\end{enumerate}

\subsubsection{``Releasing the Constraint''}

A way to mitigate the potentially aggressive nature of the fully-constrained
fit in the Sequential Empirical Bayes Method is the following: When one is
interested in estimating a particular parameter (i.e.\ the mass or weight of
the ground state or $n^{\rm th}$ excited state) one takes the priors from the
previous method and does a partially constrained fit: all other parameters are
constrained except the one in question which is unconstrained (or perhaps only
very lightly constrained).  We say that the parameter to be estimated is
``released'' (from the constraint).  The statistical error estimate obtained
this way is larger and a more conservative choice.  We will, by default,
release the constraint for all of the final results presented.

\subsubsection{``Scanning''} \label{Sect:Scanning}

Referring back to our algorithm, if a prior is available, we use it as the
value of the initial guess.  But as a new parameter is introduced there is no
prior, and an initial guess must be obtained by some other criterion.  In
practice, it may happen that different initial guesses lead to different values
for the fit parameters, especially for the unconstrained fit.  This may signal
the presence of multiple local minima in the complicated $\chi^{2}$ topography.
To address this we introduce and advocate the use of ``scanning'' over a grid
of reasonable initial values and selecting that choice which leads to a descent
to the lowest $\chi^{2}$.  By extending the scan to a rather large region of
parameter space, we ensure that the initial guesses for the parameters will not
confine the $\chi^{2}$ minimization search to a single basin of local
attraction, but rather will extend over many, including that of the global
minimum.

\subsection{Testing the Algorithm}  \label{Sect:Testing}

\subsubsection{``Reconstructing Artificial Data''}  \label{Sect:Reconstructing}

\begin{figure}[ht]
\includegraphics[angle=0,width=0.45\hsize]{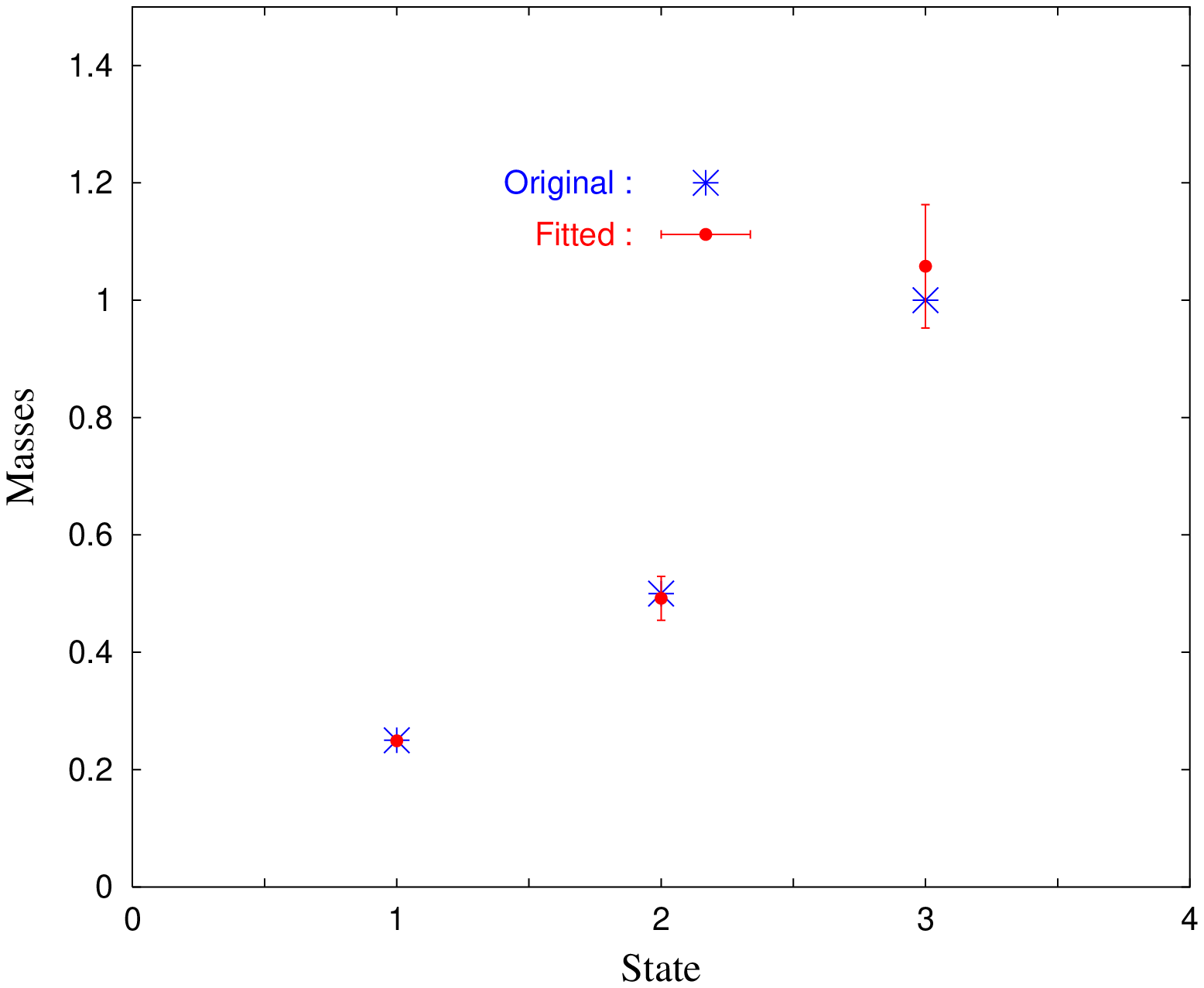}%
\includegraphics[angle=0,width=0.45\hsize]{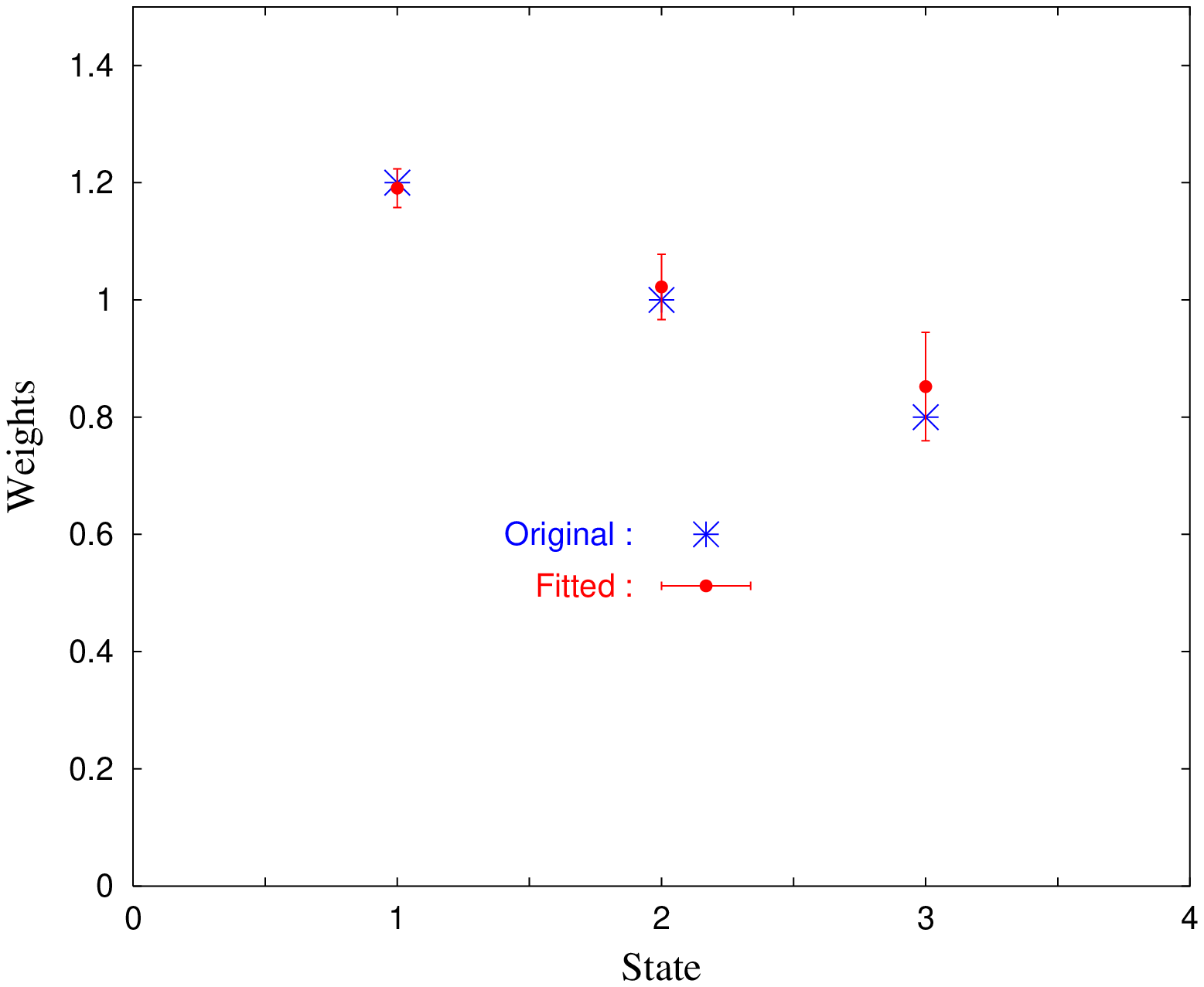}%
\caption{\label{Fig:FakeData} Recovery of parameters from artificially
constructed data.}
\end{figure}

As there are a lot of ingredients to our modification of the standard
constrained fit method (``Onion-Shell Data'', ``Scanning'', ``Global Dynamical
Weight'', ``Releasing the Constraint'') it is comforting to learn that the
method can successfully reconstruct the parameters of artificially-constructed
data where the true results are known independently of the fit.  We created a
sample of artificial data as a sum of five decaying exponentials, with means
for masses and weights fixed at values close to those extracted from real data.
Then, for simplicity, given this function, $G(t)$, we added an independent
Gaussian noise at each value of $t$.  When run through our fitting procedure,
we were able to reconstruct the masses and weights for the ground state,
first-excited state, and second-excited state (see Fig.~\ref{Fig:FakeData});
for these, the actual values were within one measured standard deviation of the
measured masses.

\subsubsection{``Partitioning the Configurations''}

\begin{figure}[ht]
\includegraphics[angle=0,width=0.5\hsize]{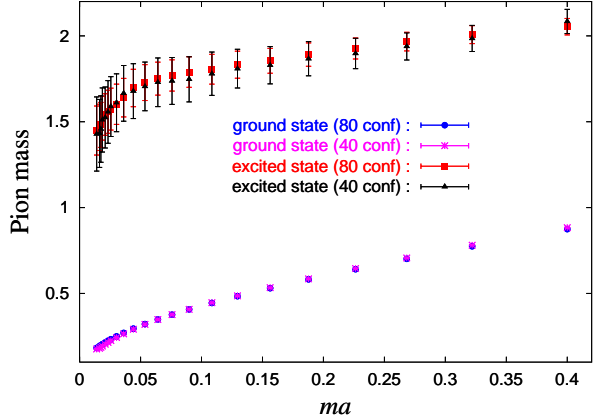}%
\caption{\label{Fig:Partition} Test of partitioning the data. Ground and first
excited state pion mass, $m_{\pi}a$, as a function of the bare quark mass
$ma$. }
\end{figure}

We have constructed an automated and natural way of obtaining the priors from a
subset of the data.  However, one worries that there is a significant amount of
``data snooping'' which may make it difficult to estimate the systematic
errors.  (Strictly speaking, using any of the data to obtain the priors
violates the Bayesian approach.)  To alleviate these worries, we have
implemented the following test: we partition the data into two non-intersecting
sets of configurations, $A$ and $B$, with an equal number, $n_{A}=n_{B}=40$ of
configurations in each set (which must still be large enough to permit stable
covariant fits).  Using the set $A$ of configurations, we perform our procedure
outlined above of obtaining the priors gradually, from the ground state up, as
the data set is monotonically enlarged by including earlier and earlier time
slices.  Now, regarding this entire procedure as a black box solely for the
purposes of obtaining the priors, we next use this fixed set of priors in the
canonical way~\cite{Lepage:2001ym} to perform a constrained fit separately on
data set $B$ (for which there is no data snooping), on data set $A$ (maximal
data snooping), and on the full set $A\cup B$ (partial data snooping but with
greater statistics).  Any disagreement beyond statistical errors can help
assess systematic errors due to data snooping.  In our case we found no
appreciable differences beyond expected statistical fluctuations.
Figure~\ref{Fig:Partition} shows a plot of the ground and first excited state
pion mass, $m_{\pi}a$, as a function of the bare quark mass $ma$.

\subsubsection{``Stability''}  \label{Sect:Stability}

\begin{figure}[ht]
\includegraphics[angle=0,width=0.5\hsize]{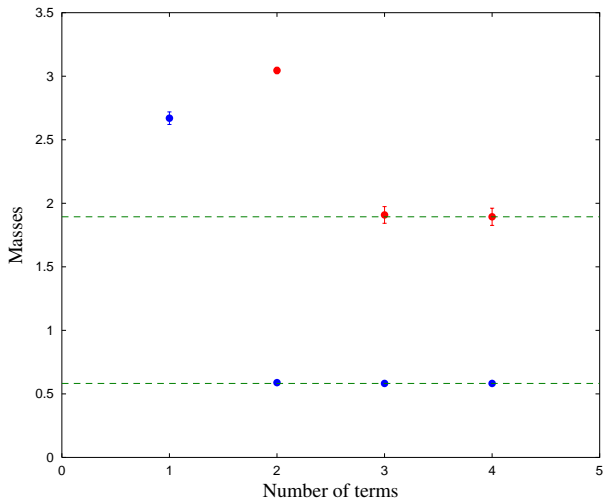}%
\caption{\label{Fig:Stability} In a test of stability, priors had previously
been selected by the Sequential Empirical Bayes Method.  Now they are used in
the standard constrained fit.  The figure shows fit values for the lowest two
masses from constrained fits (for all $t$'s) with different numbers of terms in
the fit model. The data is from the pion two-point correlation function
$\langle A_4 A_4 \rangle$ at bare quark mass $ma=0.188$ ($m_{\pi}=583(3)\,{\rm
MeV}$).}
\end{figure}

However priors are obtained, they can then be used in a standard constrained
curve fitting procedure, wherein the fit window $t_{\rm min}$--$t_{\rm max}$ is
held fixed (and as big as feasible), while the number of terms in the fit model
is increased one-by-one until the fit results converge for the lowest few
parameters of interest.  To test whether the Sequential Empirical Bayes Method
has produced reliable priors, we subsequently use its priors in the standard
way~\cite{Lepage:2001ym,Morningstar:2001je} as described above.
Figure~\ref{Fig:Stability} illustrates the passing of this test.

\subsubsection{``Fitting Errors''}  \label{Sect:FittingErrors}

\begin{figure}[hb]
\includegraphics[angle=0,width=0.5\hsize]{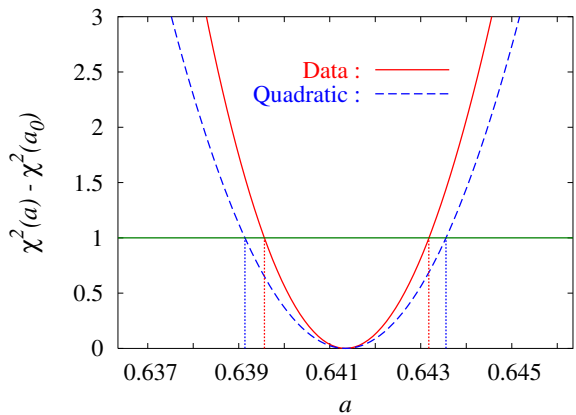}%
\caption{\label{Fig:FittingErrors} Plot of the augmented $\chi^{2}_{\rm aug}$
in the vicinity of its minimum.  It agrees well with a quadratic approximation,
ensuring that the fit error provided by the minimization routine is adequate.
The data is from the pion two-point correlation function at bare quark mass
$ma=0.226$ ($m_{\pi}=633(10)\,{\rm MeV}$).}
\end{figure}

Given the posterior probability distribution, $P(\rho|D)$, all statistics may
be evaluated by computing integrals.  In practice, such evaluation is often
difficult to perform directly; Monte Carlo methods such as simulated
annealing~\cite{Fiebig:2002sp} can be used.  Even so, the computational cost is
often daunting and so approximations are sought.  Most commonly, one may assume
that the data set is sufficiently large that the Central Limit Theorem implies
that $\chi^{2}_{\rm aug}$ is approximately quadratic about its minimum.  We in
fact use this approximation, and furthermore use the resulting fitting errors
from a fit as the priors $\tilde{\sigma}$ for the next fit.  To test this, we
plot in Fig.~\ref{Fig:FittingErrors} the $\chi^{2}_{\rm aug}$ as a function of
$\rho$ in the neighborhood of the minimum.  Superimposed is the parabola which
is the quadratic approximation about the minimum.  The two agree very well.
Also calculated and plotted as ranges are the ``fit error'', obtained from the
quadratic approximation, and the second moment of the actual distribution.
These too agree very well, and indicate that using the naive fit error as
provided by the minimization routine is quite adequate.

Although fit errors are used to produce priors for the inner loops of the
algorithm, all final errors reported are bootstrap errors.

\subsubsection{Caveat Emptor: Further Toy Model Results}  \label{Sect:Caveat}

It is crucial to be extremely conservative when making fits, especially when
relying on Bayes' inference methods.  One must be sure to avoid announcing a
``false positive'', that is making claim which cannot be guaranteed to be
correct.  We can understand in more detail how a false positive may emerge and
how to design the procedure to avoid them, by testing variants of the SEB
algorithm against a number of ``toy models'' where the actual values of the
function's parameters are known.

Suppose the artificial data is created with the three-state model function
\begin{eqnarray}
      G^{\rm true}(t) 
& = & 
      w_1 e^{-m_1 t} 
    + w_2 e^{-m_2 t} 
    + w_3 e^{-m_3 t} \\ \nonumber
& = & 
      w_1 e^{-m_1 t}
      \left(1+\frac{w_2}{w_1}e^{-(m_2-m_1)t} 
             +\frac{w_3}{w_1}e^{-(m_3-m_1)t}\right),
\end{eqnarray}
and we want to perform a fit of the data sample generated by this function with
fixed values of input parameters and Gaussian statistical error $\delta G(t)$
at each time slice $t$.  Then SEB should work if the following holds:
\begin{enumerate}
  \item There exists a $t_1$ such that in the time range $t_1<t<t_{\rm max}$, 
  \begin{equation}
    \frac{w_2}{w_1}e^{-(m_2-m_1)t} 
    <  
    \delta G(t)\frac{1}{G(t)},
  \end{equation}
   so that in this range, the data can be fitted by $w_1 e^{-m_1 t}$.  That is,
   there is a ``plateau'' in the effective mass plot for large time.
  \item There exists a $t_2$ such that in the time range $t_2 \le t < t_1$,
  \begin{equation}
    \frac{w_3}{w_1}e^{-(m_3-m_1)t} 
    < 
    \delta G(t)\frac{1}{G(t)}<\frac{w_2}{w_1}e^{-(m_2-m_1)t}
  \end{equation}
  so that in this range, the data can be well fitted by $w_1 e^{-m_1 t} + w_2
  e^{-m_2 t}$.
  \item In the time range $t < t_2$, the third state becomes important, and the
  full three-state model must be used in the fit.
\end{enumerate}

Now the SEB is very adaptive; in the course of the iterative procedure, the
ground state is fitted several times and allowed to float within a sigma or so
of its current prior with each new fit.  Thus the original criterion of the
existence of a clear plateau need not be strictly enforced.  But if the
condition for the plateau is badly violated, or if the similar conditions for
the excited states are badly violated, there is a possible danger of the
algorithm gravitating toward a local minimum in the $\chi^2$ which is not the
true value.

To illustrate, consider artificial data, constructed from the following
three-state toy model
\begin{eqnarray} \label{ToyModel}
      G(t;w_i,m_i) 
& = & 
      \sum_{i=1}^{3} w_i e^{-m_i t} \\ \nonumber
& = & 
      500 e^{-0.85 t} + 400 e^{-1.35 t} + 400 e^{-1.75t}
\end{eqnarray}
in the time range $0\le t \le 15$, with relative errors (uncorrelated and
Gaussian) increasing with time $t$ to mimic actual LGT data.

In keeping with the spirit of the SEB, we fit first to a single exponential
over the time interval $t \in [t_1,t_{\rm max}]$, then use the fitted ground
state values as priors for a two-state partially-constrained fit in the time
interval $t \in [t_2,t_{\rm max}]$, and then use the fitted ground and
first-excited state fit values as priors for a final three-state
partially-constrained fit in the time interval $t \in [t_{\rm min},t_{\rm
max}]$, where $t_{\min}=0 \le t_2 \le t_1 \le t_{\rm max}=15$.  But to stress
the caveat about the dangers of ``false positives'', we explore all values for
the pair $(t_1,t_2)$.

\paragraph{``Lowest Precision'' case:} With statistical errors ranging from $\delta
G/G\sim 0.011$ at $t=0$, to $\sim 0.12$ at $ t= 15 $, the lowest $\chi^2/{\rm
dof}$ fit is for $(t_1,t_2)=(10,7)$ with $m_1=0.855(7)$, $w_1=521(34)$,
$m_2=3.10(62)$, $w_2=0.0004(1)$, $m_3=1.51(5)$, $w_3=769(35)$, that is, the fit
fails to produce the input parameters of the toy model.

\paragraph{``Low Precision'' case:} If the Gaussian noise is reduced by about a factor of two 
to $\delta G(t)/G(t) \sim 0.006$ at $t=0$, $\delta G(t)/G(t) \sim 0.05$ at $ t=
15 $, then there are two solutions with the lowest $\chi^{2}/{\rm dof}=0.527$:
a false positive with $(t_1,t_2)=(10,2)$ and $m_1=0.854(4)$, $w_1=518(23)$,
$m_2=1.50(8)$, $w_2=701(63)$, $m_3=1.86(92)$, $w_3=69(62)$, and a fit which
agrees with the input model: $(t_1,t_2)=(10,4)$ with $m_1=0.853(5)$,
$w_1=511(29)$, $m_2=1.39(16)$, $w_2=369(188)$, $m_3=1.65(17)$, $w_3=408(188)$.

\paragraph{``High Precision'' case:} If the Gaussian noise is further reduced to 
$\delta G(t)/G(t) \sim 0.001$ at $t=0$, $\delta G(t)/G(t) \sim 0.01$ at $t= 15$
then the lowest $\chi^2/{\rm dof}=0.530$ does reproduce the input parameters:
$(t_1,t_2)=(10,5)$, $m_1=0.851(1)$, $w_1=504(7)$, $m_2=1.38(6)$,
$w_2=451(141)$, $m_3=1.77(10)$, $w_3=343(144)$.

Let $\Delta G(t)$ be the absolute value of the difference of the function
$G^{\rm true}(t)$ of input parameters and the function $G^{\rm fitted}(t)$ of
fitted parameters
\begin{equation}
\Delta G(t) = |G^{\rm true}(t)-G^{\rm fitted}(t)|.
\end{equation} 
A fit $G^{\rm fitted}(t)$ with reasonable small $\chi^2/{\rm dof}$ implies that
the relation
\begin{equation}  \label{DeltaG}
  \Delta G(t) 
  < 
  \delta G(t)
\end{equation} 
roughly holds at most times $t$.  In other words, with this statistical error
$\delta G(t) $, we cannot distinguish the fitted function $G^{\rm fitted}(t)$
from the original $G^{\rm true}(t)$.

\begin{figure}[ht]
\includegraphics[angle=0,width=0.7\hsize]{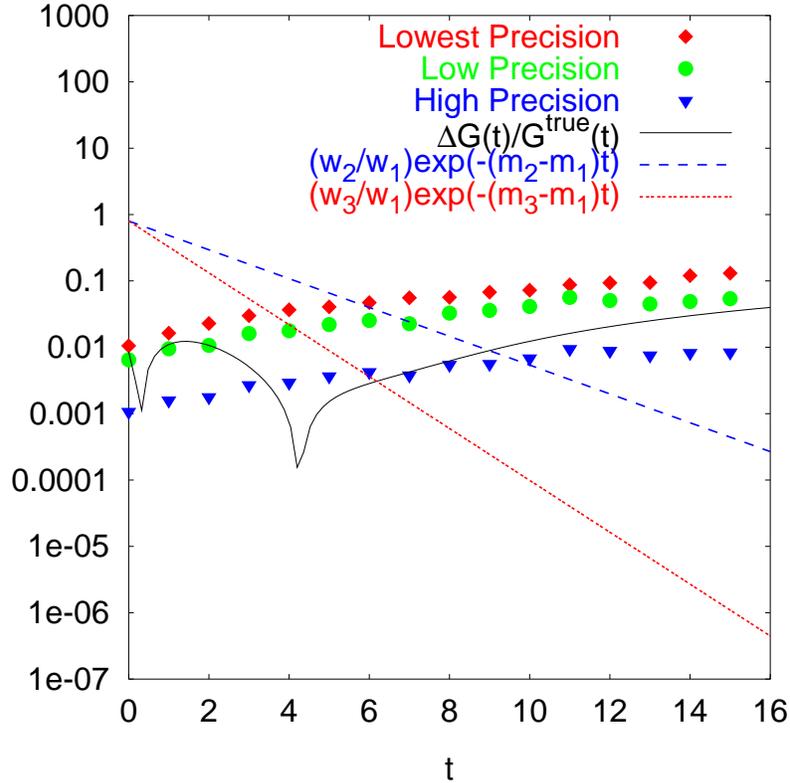}%
\caption{\label{Fig:RelativeError} The relative errors of three data sets of
increasing precision (``Lowest'', ``Low'', and ``High'') are plotted with
points.  The curved solid (black) curve is the plot of the function $\Delta
G(t)/G^{\rm true}(t)$, which is the relative difference between original
function $G^{\rm true}(t) = 500 \exp(-0.85 t) + 400 \exp(-1.35 t) + 400
\exp(-1.75 t)$ and the false-positive fitted function $G^{\rm false} = 521
e^{-0.855 t} + 769 e^{-1.51 t}$ from the data set (``Lowest Precision'').  }
\end{figure}

From Fig.~\ref{Fig:RelativeError} we see that the statistical errors of the
data set (``Lowest Precision'') are all larger than the difference of the two
functions in the full time range.  This is the reason why the false positive
can be chosen by the fit routine.  The dashed (blue) straight line ($w_2/w_1
\exp(-(m_2-m_1)t)=0.8\exp(-0.5t)$) and the dotted (red) straight line ($w_3/w_1
\exp(-(m_3-m_1)t)=0.8\exp(-0.9t)$) intersect with the curve of relative errors
at $t_1$ and $t_2$, respectively.  This shows that the ground state dominates
in the range $t\ge t_1$, the first-excited state plays a role in the range
$t_2\le t<t_1$, and the three states should all be included in the time range
$t<t_2$.  When the precision is higher, the statistical errors are smaller than
the difference of two functions, in some time range, so that the fit can reject
the false positive $G^{\rm false}(t)$.

Returning to the false positive of the ``Lowest Precision'' case, notice that
when forced to fit to a three-state model, the fit prefers a two-state solution
with the one weight essentially zero, and the remaining weight consistent with
the sum of the true second and third weights and the remaining mass
interpolating between the true masses.  In short, the fit is content with
averaging the two higher states into one.  Also notice that it is the second
weight which is zero.  This is because, at the intermediate stage, a two-state
model was forcing a fit to data completely dominated by a single state.
Subsequently, as earlier times were added, the data could be described by two
states, but since the would-be second state weight was by now constrained near
zero, its role was supplanted by the third state in the model.

That is, by introducing the new (second) state into the model before the data
was precise enough to discern it, spurious results were obtained.  This
suggests the cure: postpone the introduction of new states in the model until
the data demands it (as evidenced by a sudden increase in the $\chi^2/{\rm
dof}$).

In more detail, if the time range is large enough there exists a largest time
$t_1$ beyond which the contamination of higher states will be obscured by the
statistical errors and can be neglected so that in the time range $t\ge t_1$
the ground state dominates and the data can be well described by the single
exponential $w_1 \exp(-m_1 t)$.  Thus the time range $t\ge t_1$ is a good
choice for the first step of SEB, and we can get a reasonably small
$\chi^2/{\rm dof}$ over this ground-state effective-mass plateau.  If we
include more time slices with $t_2 \le t < t_1$, the contribution of the first
excited state becomes significant, but the second excited state contributes
little to the correlation function in this range.  The presence of the first
excited state results in a noticeable increase of the $\chi^2/{\rm dof}$ if we
continue to force a single exponential as a fit model; the inclusion of one
more term in the fit model is necessary.  Continue this procedure until the
time-slices are exhausted.

\bigskip

{\em The Penultimate Algorithm: ``Not-too-Soon''}
\begin{enumerate}

  \item The available data are in the time range $t_{min}-t_{max}$.

  \item From effective mass plots, choose an initial time range $t\in[t_{\rm
  start},t_{\rm max}]$ to do an unconstrained one-mass fit.  Use ``scanning''
  -- see Sect.~\ref{Sect:Scanning}.

  \item Include one more time slice and repeat an (independent) unconstrained
  one-mass fit, and monitor the fitted parameters ${m_1, w_1}$ and the
  $\chi^2/{\rm dof}$.

  \item If the fitted parameters and the $\chi^2/{\rm dof}$ do not change much,
  include one more time slice and repeat the previous step.  This iteration
  stops if there is a noticeable change of $\chi^2/{\rm dof}$ and the values of
  the fitted parameters indicating a breakdown of the one-state model.  Then
  set $t_1-1$ equal to the time at which the $\chi^2/{\rm dof}$ jumps,
  indicating the necessity of a two-mass fit for $t<t_1$.  Set the priors for
  the ground state mass and weight equal to the fitted values from the last
  low--$\chi^2/{\rm dof}$ fit over $t\in[t_{1},t_{\rm max}]$.

  \item Include one more mass term in the fit model and do a
  partially-constrained two-mass fit (with scanning) for $t\in[t_{1}-1,t_{\rm
  max}]$.  The ground state priors are fixed at the values determined by the
  previous step.  The first-excited state is unconstrained but scanned.

  \item Repeat, adding time slices until the two-state model breaks down as
  indicated by a jump in the $\chi^2/{\rm dof}$.  Then set $t_2-1$ equal to the
  time at which the $\chi^2/{\rm dof}$ jumps, indicating the necessity of a
  three-mass fit for $t<t_2$.  Set the priors for the ground state and
  first-excited state mass and weight equal to the fitted values from the last
  low--$\chi^2/{\rm dof}$ fit over $t\in[t_{2},t_{\rm max}]$.  Note that the
  ground-state priors are refreshed.

  \item Repeat, adding more time slices and more fit-model terms (one at a time
  and only when necessary) and more until all time slices are used.

  \item The highest state in the fit model will be absorbing all the
  contributions from higher states in the true function, and thus its fitted
  parameters will differ from the true values.  Thus the highest state in the
  fit model must be rejected.

\end{enumerate}

Now we return to the artificial data from the toy model ``Lowest Precision''.
Recall that when we assume a three-mass fit and independently try all pairs
$(t_1,t_2)$, we obtained a false positive.  But the statistical errors are so
large that they obscure the contributions from excited states at all but the
earliest time slices.  Adding terms to the fit model prematurely led to
spurious results.  Now we apply the new method outlined immediately above to
the same data.

A one-state fit works well for $t\in[t_1,t_{\rm max}=15]$ until the $\chi^2$
suddenly jumps at the (proposed value of) $t_1=3$.  Thus we set $t_1=4$.  The
values of the one-mass term fit over time range $[4,15]$ are used to set priors
for $m_1$ and $w_1$, and a second term is added to the fit model.  The two-term
model works well for a fit in the range $[t_2,15]$ with $t_2$ varying from 3 to
0.  The two-mass model fit the data very well ($\chi^2/{\rm dof}=0.265$) in the
whole time range with fitted parameters $m_1 = 0.855(7)$, $w_1 = 521(34)$, $m_2
= 1.51(5)$, $w_2 = 768(35)$.  Since the two-mass fit exhausts all the time
slices and does not show a noticeable increase of $\chi^2/{\rm dof}$, we don't
go ahead to include the third mass term in the fit model.

But this fit is a ``false positive''!  Since the statistical errors are larger
than the difference between the false positive and the true solution
(Eq.~\ref{DeltaG}), the false positive cannot be rejected on the basis of its
$\chi^2/{\rm dof}$, by this or any other method.  So what have we gained?  The
difference this time is that this solution is exposed as a two-mass-term fit
rather than a spurious three-term fit.  And as always, the highest state in the
fit must routinely be dropped since in general it can be contaminated by
contributions from higher excited states.

We conclude that with the ``Lowest Precision'' data set, we cannot get an
unambiguous estimate of the second and the third state.  The fit is comfortable
predicting the ground state parameters only (and they are correct).  Thus the
algorithm has not made a ``false positive'' claim.

\begin{figure}[ht]
\includegraphics[angle=0,width=0.45\hsize]{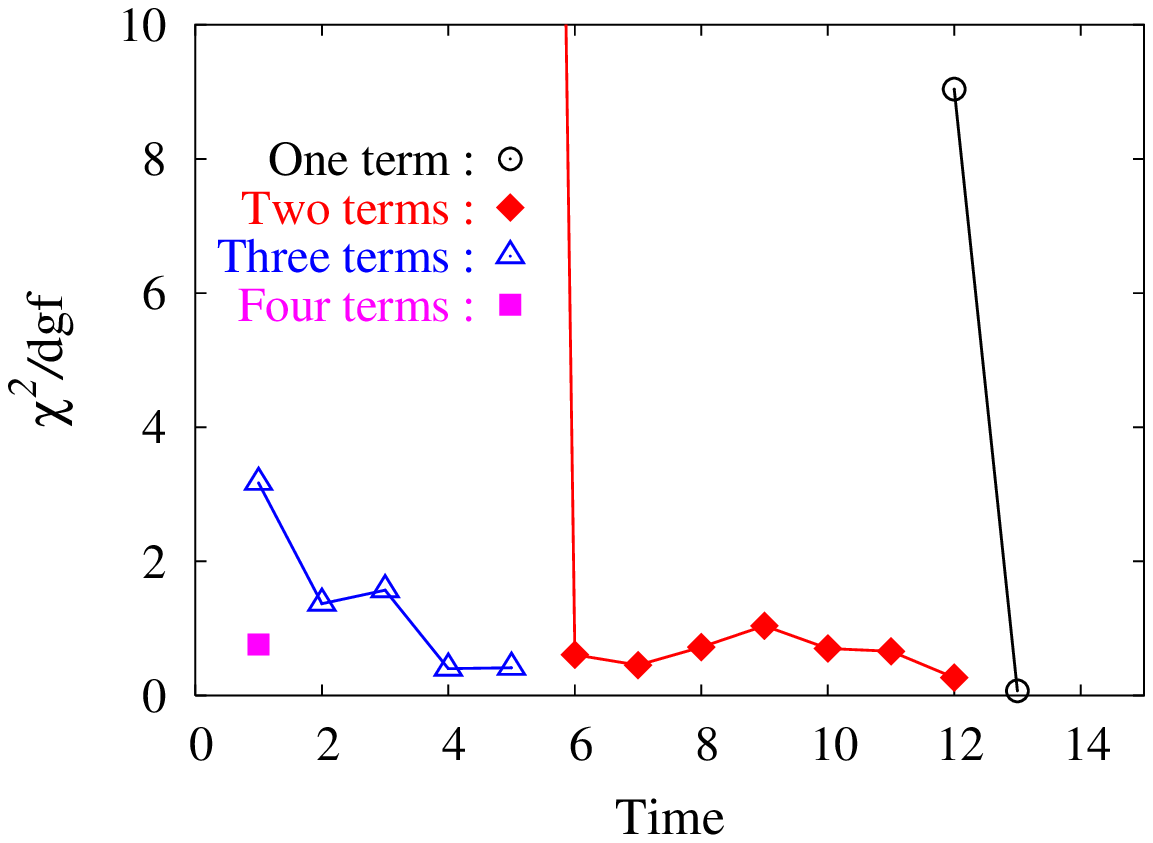}%
\includegraphics[angle=0,width=0.45\hsize]{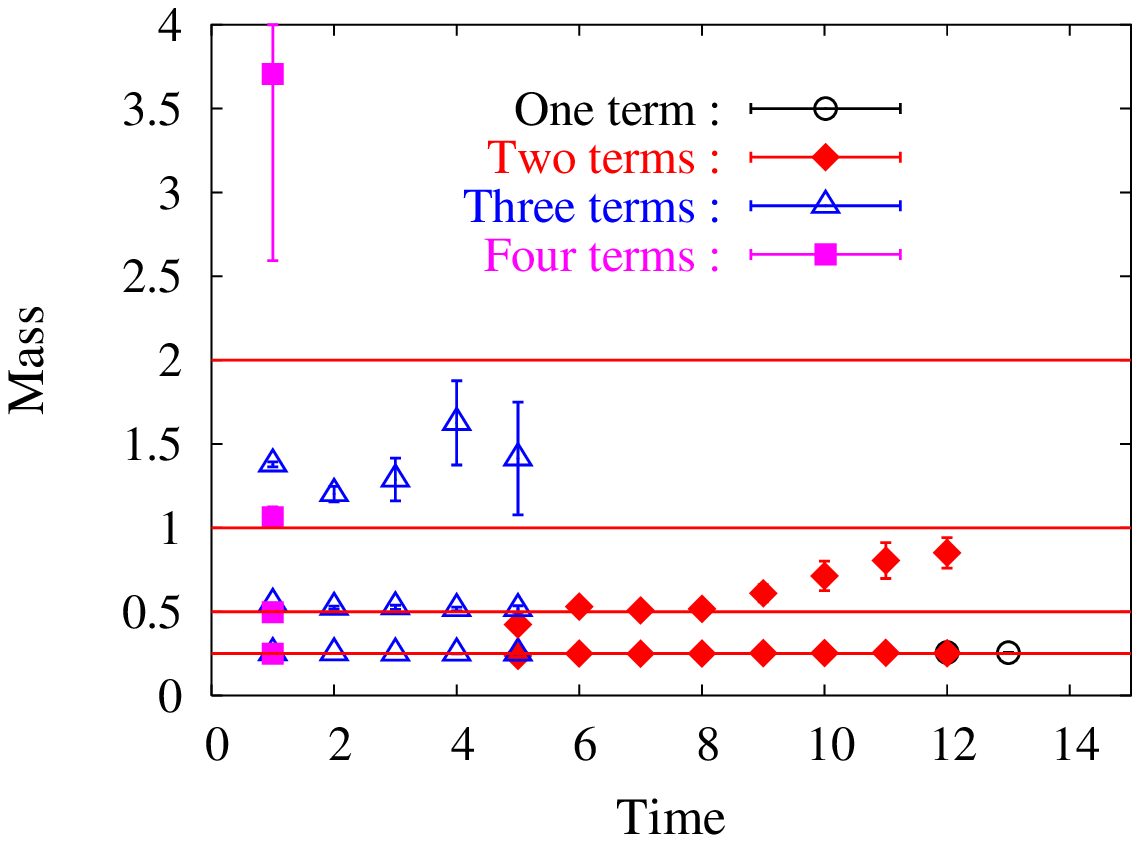}%
\caption{\label{Fig:ToyChiSq1} Behavior of the $\chi^2/{\rm dof}$ of the fit as
earlier time slices are added to the fit range.  A sudden jump indicates that
the fit model is inadequate and that a new mass term should be added to the
model.  Accordingly, then the $\chi^2/{\rm dof}$ drops only to increase again
as further time slices are added.  ``Not-too-Soon'' addition of terms to the
fit model prevents a single state being erroneously identified as two.  Also
shown are the behavior of the ground and first few excited states, which
stabilize as more terms are added.   }
\end{figure}

Now, we illustrate this technique with another set of artificial data for which
it will be possible to extract excited states.  It will be instructive to
monitor at the behavior of the $\chi^2/{\rm dof}$, but also to see how the fit
parameters of each term in the fit model stabilize as time slices are added.
Fig.~\ref{Fig:ToyChiSq1} (left) plots the $\chi^2/{\rm dof}$ for data
artificially constructed as a sum of four decaying exponentials.  A one-term
fit is adequate for time slice 13 ($t_{\rm max}=15$).  But it is exposed as
being entirely inadequate as time slice 12 is added to the data set to be fit.
(The $\chi^2/{\rm dof}$ jumps tremendously because we have made the statistical
errors on the data to be fit quite small, to illustrate the point.)  So for
time slices 6-12 a two-term fit model is used, and the fit is quite adequate.
At time slice 5, however, the $\chi^2/{\rm dof}$ jumps tremendously, exposing
the two-term fit as inadequate.  Thus a third term is added at time slice 5.
The three-term fit works down through time slice 2.  At time slice 1, a fourth
term must be added to keep the $\chi^2/{\rm dof}$ from being unreasonably
large.

Meanwhile, Fig.~\ref{Fig:ToyChiSq1} (right) monitors the masses of the lowest
few states as more time slices are added to the data set and more terms are
added (as necessary) to the fit model.  At time slice 13, the one-term model is
fitted with a ground-state mass (open circle).  (Its value is consistent with
the value obtained much later when time slices 1--12 have been added to the fit
model and three more terms are added to the fit model.  That is, time slices
above 13 are sufficient to determine the ground-state mass, i.e the effective
mass plot has ``plateaued''.  But more importantly for the efficacy of the
algorithm, this fitted value remains stable later as the data and fit model are
expanded.)  At time slice 12, the second term is added to the fit model (solid
diamonds).  As each earlier time is added, the first-excited state mass changes
by a sigma or so, as the augmented data is refit.  This is important.  It is
saying that if a relatively small amount of data from time slices 12 and above
suggested a somewhat inaccurate guess for the prior for this mass, then
subsequent fits including data at time slices 8-11 can influence the value and
bring it toward the correct value (at the horizontal line).  By time slice 8,
the first excited state mass has stabilized (at the correct value) and will not
subsequently deviate through time slice 6.  At time slice 5, the first-excited
state mass deviates from the correct answer.  But this is where the jump in the
$\chi^2/{\rm dof}$ warns that the two-state fit model is inadequate; the
first-excited state fitted mass is contaminated from higher states in the data.
Adding a third term returns the fitted first-excited state mass to its correct
value.  It remains stably at this value with the addition of earlier time
slices 2--4 to the three-state fit model (open triangles).  Before the
second-excited state has stabilized, we have run out of earlier time slices to
add, and we can make no claim as to whether its fitted value is correct.  (It
isn't.)  But the algorithm has successfully fitted the first-excited state (and
of course the ground state).

\begin{figure}[ht]
\includegraphics[angle=0,width=0.45\hsize]{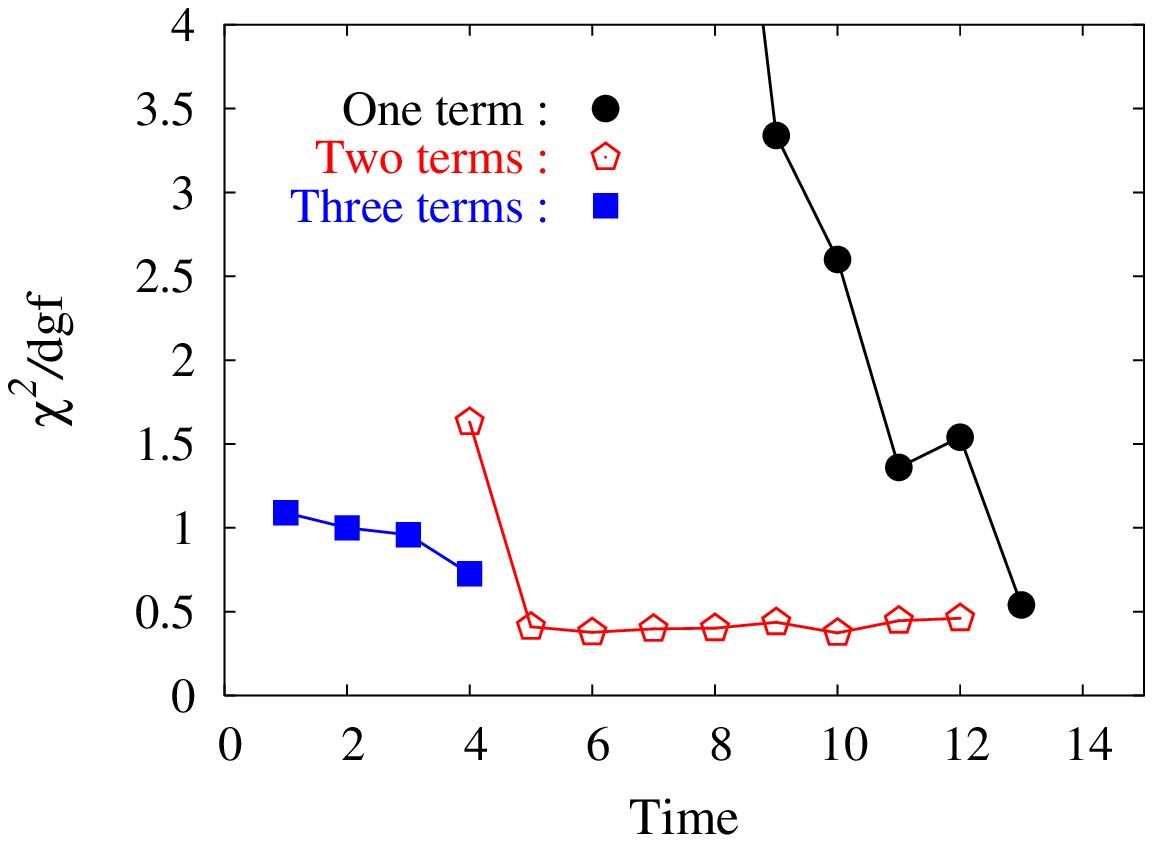}%
\includegraphics[angle=0,width=0.45\hsize]{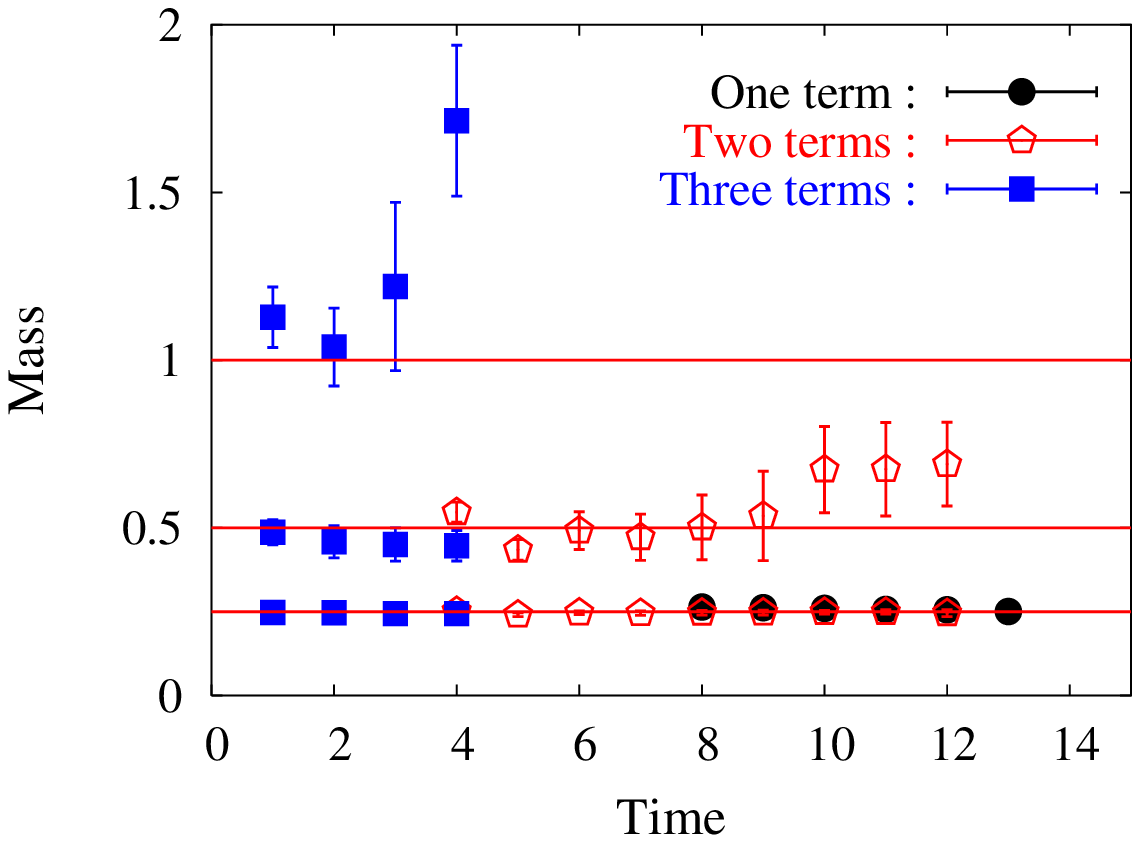}%
\caption{\label{Fig:ToyChiSq2} Same as in Fig.~\ref{Fig:ToyChiSq1} but the
artificial data is less precise (1\%).  Accordingly, the $\chi^2/{\rm dof}$
does not jump to such extreme values when the number of the terms in the fit
model becomes inadequate.}
\end{figure}

\begin{figure}[ht]
\includegraphics[angle=0,width=0.45\hsize]{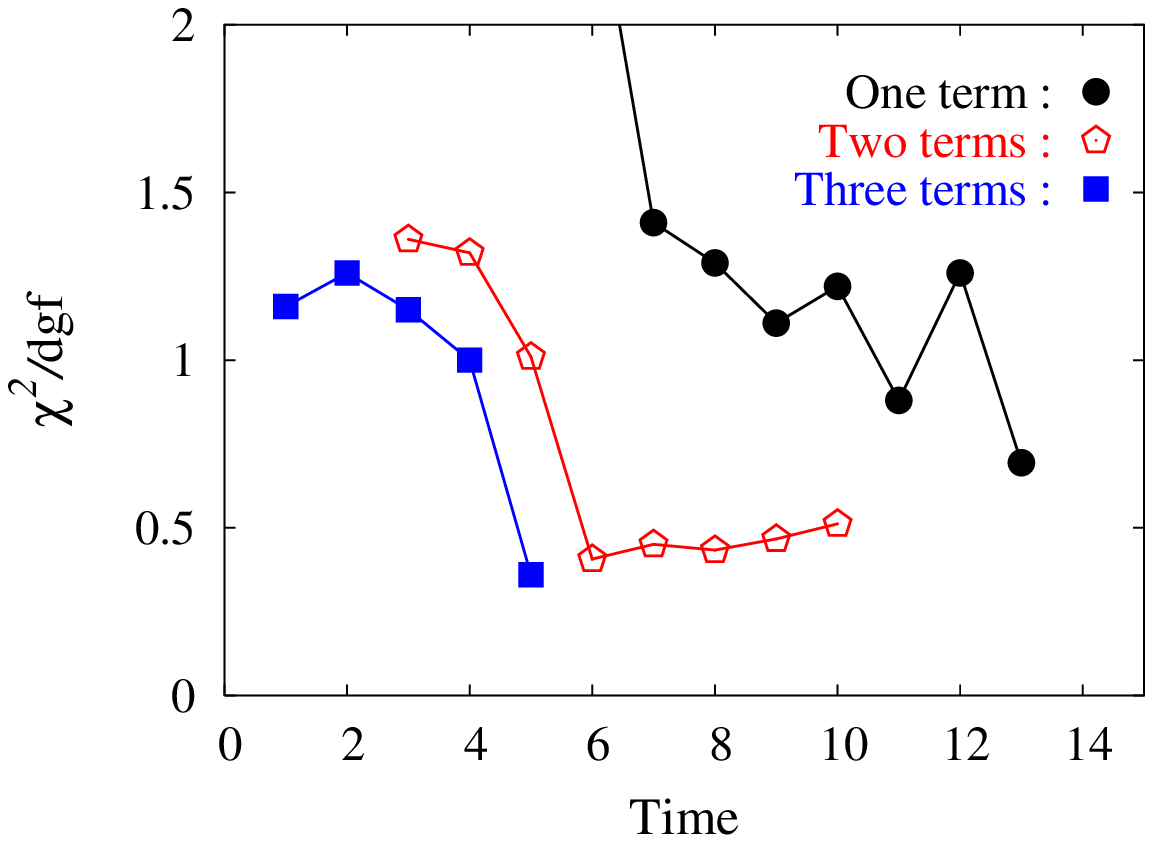}%
\includegraphics[angle=0,width=0.45\hsize]{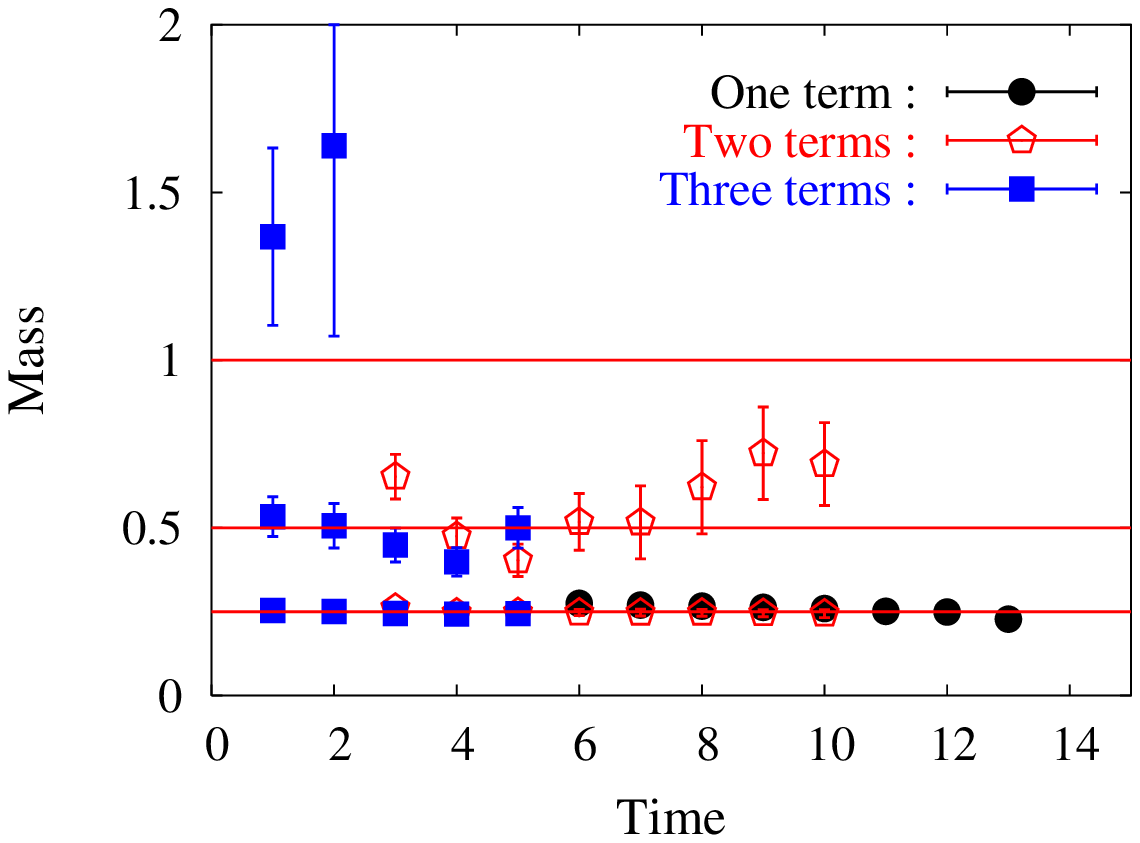}%
\caption{\label{Fig:ToyChiSq3} Same as in Fig.~\ref{Fig:ToyChiSq1} but the
artificial data is even less precise (5\%).}
\end{figure}

Figs.~\ref{Fig:ToyChiSq2} and~\ref{Fig:ToyChiSq3} tell a similar story for
artificial data which is less precise (1\% and 5\% respectively).  The
$\chi^2/{\rm dof}$ jump (indicating that another term should be added to the
fit model) but, as one would expect, not as extremely as for the more precise
data of Figs.~\ref{Fig:ToyChiSq1}.

\subsubsection{Our Final Algorithm: ``Just-in-Time''}  \label{Sect:FinalAlgorithm}

As we saw in the last section, the ``Not-too-Soon'' algorithm avoids ``false
positives'' by delaying the addition of a new term to the fit model before the
data were able to discern it.  However, there is the logical possibility of the
opposite danger: if the new term is added too late, then there could
potentially be an erroneous $n$-state fit of data which is in fact better
described by $n+1$ states.  If so, then the highest two states will be
``averaged'' into one.  (So, for example, a fitted first excited state mass
might take a value intermediate between the true first and second excited state
masses.)  The addition of further states to the fit model might then cover up
this misidentification, and allow for this different kind of false positive.

One would think that an increased $\chi^2/{\rm dof}$ would warn of this danger;
however, since the $\chi^2/{\rm dof}$ is an average over many time slices, it's
warning may come a time slice or two too late.  If the data is such that a new
term is discernible every couple of slices, then this may indeed lead to
erroneous results.

\begin{figure}[ht]
\includegraphics[angle=0,width=0.45\hsize]{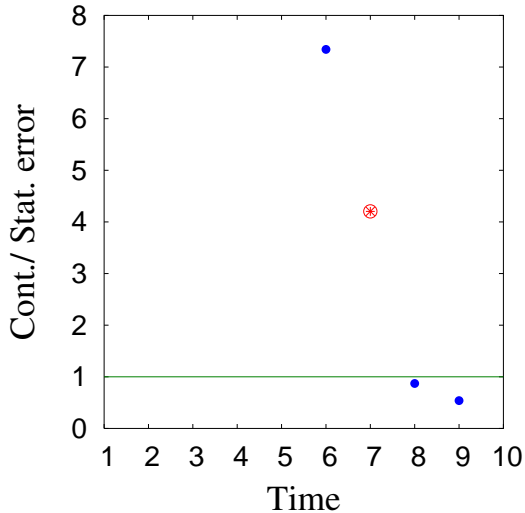}%
\caption{\label{Fig:RelativeContribution} Ratio of the contribution to the fit
model from a proposed new term, $w_{n}e^{-m_n t_i}$, compared to the
statistical error $\delta G(t_i$) at the newest time slice $t_i$ added to the
data set.  The new term is deemed not needed at time slices $t_i=9$ and
$t_i=8$.  At time slice $t_i=7$, the ratio exceeds $1$ and the new term is
conditionally accepted.  The ratio remains above $1$ for time slice $t_i=6$ and
below, confirming that the new term is needed.  The data is from a fit of the
Roper resonance at the same quark mass as in
Fig.~\ref{Fig:Roper_stability_previous}.}
\end{figure}

A final tweak of the algorithm addresses this issue: The ``Just-in-Time''
algorithm is the same as the ``Not-too-Soon'' algorithm outlined in the last
section, except that the previous criterion for adding a new term to the fit
model, namely that without this new term the $\chi^2/{\rm dof}$ would suddenly
increase, is replaced by a more sensitive criterion: As a new time slice $t_i$
is added to the data set, a tentative fit is made with a proposed extra term,
\{$w_n$,$m_n$\}.  Then the contribution of this fitted term is compared to the
statistical error at the new time slice
\begin{equation}\label{AcceptCriterion}
w_{n}e^{-m_n t_i} > \delta G(t_i) 
\end{equation}
(See Fig.~\ref{Fig:RelativeContribution}.)  If Eq.~\ref{AcceptCriterion} does
not hold, then the new term is deemed not needed.  If it does hold, then the
new term is accepted (but only conditionally since the equation may hold only
by statistical fluctuation).  As new time slices are added, the new term must
continue to pass this test.  (This is stable: typically, if it passes the test
once, it is likely to continue to pass the test at earlier time slices, since
for these the relative error gets smaller and the fractional contribution gets
larger.)  If the test is failed before a higher-order term is conditionally
added, then one returns to time slice $t_i$ and it is deemed that the $n$th
term is not included in the fit.  Then time $t_i - 1$ is added to the data, a
new proposal is made to add the $n$th term, and the process continues.

\section{Some Physics Results} \label{Sect:Results}

We have presented a list of algorithms with steadily increasing complexity but
with steadily increasing robustness.  The earliest and simplest ``fixed $\Delta
t$'' algorithm, most easily described and presented first here as a template,
was originally used (prematurely in retrospect) for some preliminary conference
presentations.  We now deem it too unsophisticated and do not use it anymore.
The plain vanilla ``variable $\Delta t$'' algorithm is adequate provided that
the level spacings are well separated and each term is saturated by several
time slices.  It is perfectly adequate to describe, for instance, the pion
ground and excited state, as explained in Sect.~\ref{Sect:Pion}.  The plain
``variable $\Delta t$'' was used in the first version of our Roper paper and
caused some erroneous results at high quark mass.  This has been rectified by
the use of the ``Just-in-Time'' algorithm for the updated version of our Roper
paper and is briefly described in Sect.~\ref{Sect:Roper}.

\subsection{The Simulation} \label{Sect:Simulation}

On a $16^{3}\times 28$ lattice, we use the overlap
fermion~\cite{Neuberger:1998fp} and the Iwasaki gauge
action~\cite{Iwasaki:1985we} with $\beta=2.264$.  The lattice spacing,
determined from the measured pion decay constant $f_{\pi}$, is determined to be
$a=0.200(3)\,{\rm fm}$, and thus the lattice has spatial size of $3.2\,{\rm
fm}$.

We adopt the following form for the massive Dirac
operator~\cite{Hernandez:1999cu,Alexandrou:2000rg,Capitani:2000wi}
\begin{equation}  \label{neu}
D(m_0)= (1 - \frac{m_0a}{2\rho})\rho D(\rho) + m_0a,
\end{equation}
where
\begin{equation}
D(\rho) = 1 + \gamma_5 \epsilon (H),
\end{equation}
so that
\begin{equation}
D(m_0) = \rho + \frac{m_0a}{2} + (\rho - \frac{m_0a}{2} ) \gamma_5 \epsilon (H),
\end{equation}
where $\epsilon (H) = H /\sqrt{H^2}$ is the matrix sign function and $H$ is
taken to be the hermitian Wilson-Dirac operator, i.e. $H = \gamma_5 D_w$.  Here
$D_w$ is the usual Wilson fermion operator, except with a negative mass
parameter $- \rho = 1/2\kappa -4$ in which $\kappa_c < \kappa < 0.25$. We take
$\kappa = 0.19$ in our calculation which corresponds to $\rho = 1.368$. The
massive overlap action is defined so that the tree-level renormalization of
mass and wavefunction is unity.

We adopt the Zolatarev
implementation~\cite{vandenEshof:2001hp,vandenEshof:2002ms} of the optimal
rational approximation~\cite{Edwards:1998wx,Dong:2000mr} to approximate the
matrix sign function. The inversion of the quark matrix involves nested do
loops in this approximation.  Further details of the procedure are given
elsewhere~\cite{Dong:2001fm,Dong:2000mr}.

\subsection{Pion} \label{Sect:Pion}

\begin{figure}[ht]
\includegraphics[angle=0,width=0.5\hsize]{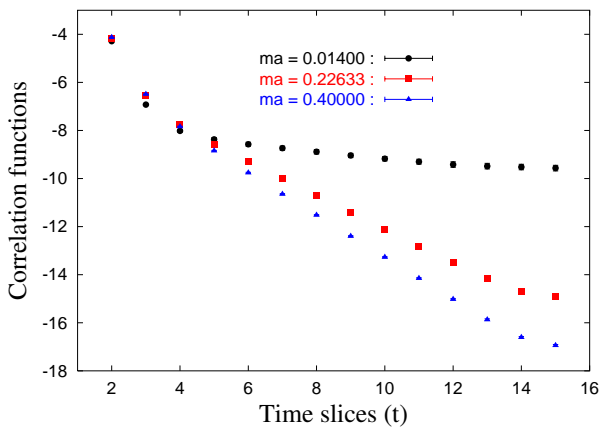}%
\caption{\label{Fig:PiCor} Two-point correlation function $\langle A_4 A_4
\rangle$ for the pion for three bare quark masses.}
\end{figure}

Figure~\ref{Fig:PiCor} shows the correlation function for the correlator
$\langle A_4 A_4 \rangle$.  One can see that it is dominated by the ground
state of the pseudoscalar channel (pion) over all but the few earliest time
slices.  This presents a problem for the default ``fixed $\Delta t$'' approach.
Referring back to the algorithm of section~\ref{Sect:BasicAlgorithm}, we see
that at each step of the algorithm, $\Delta t$ new time slices are added to the
data while a new excited state is added to the fit model.  Thus for the pion
correlator, one would be trying to fit to a model with many states when the
data is saturated by just the ground state.  Forcing a fit may give misleading
estimates of excited state parameters which are subsequently used as priors.

With the ``variable $\Delta t$'' refinement of the algorithm, rather than
deciding {\it a priori\/} on the number of terms in the fit and adding time
slices a fixed number at a time, one lets the data decide how many time slices
to include with each enlargement of the data by choosing the minimum $\chi^{2}$
over a range of reasonable possibilities.  Thus since the pion correlator is
dominated by the ground state for many time slices, then many time slices will
be automatically added before an attempt is made to fit the first-excited
state.

\begin{figure}[ht]
\includegraphics[angle=0,width=0.5\hsize]{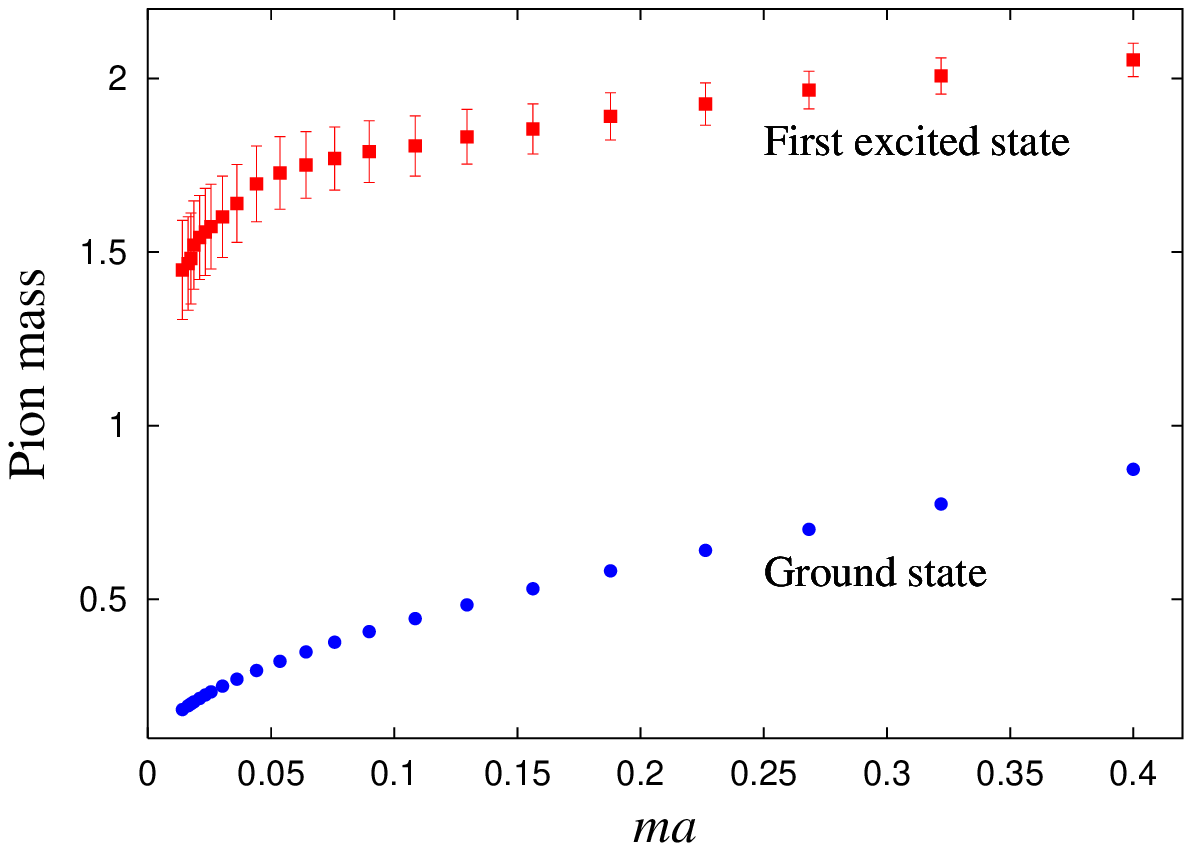}%
\includegraphics[angle=0,width=0.5\hsize]{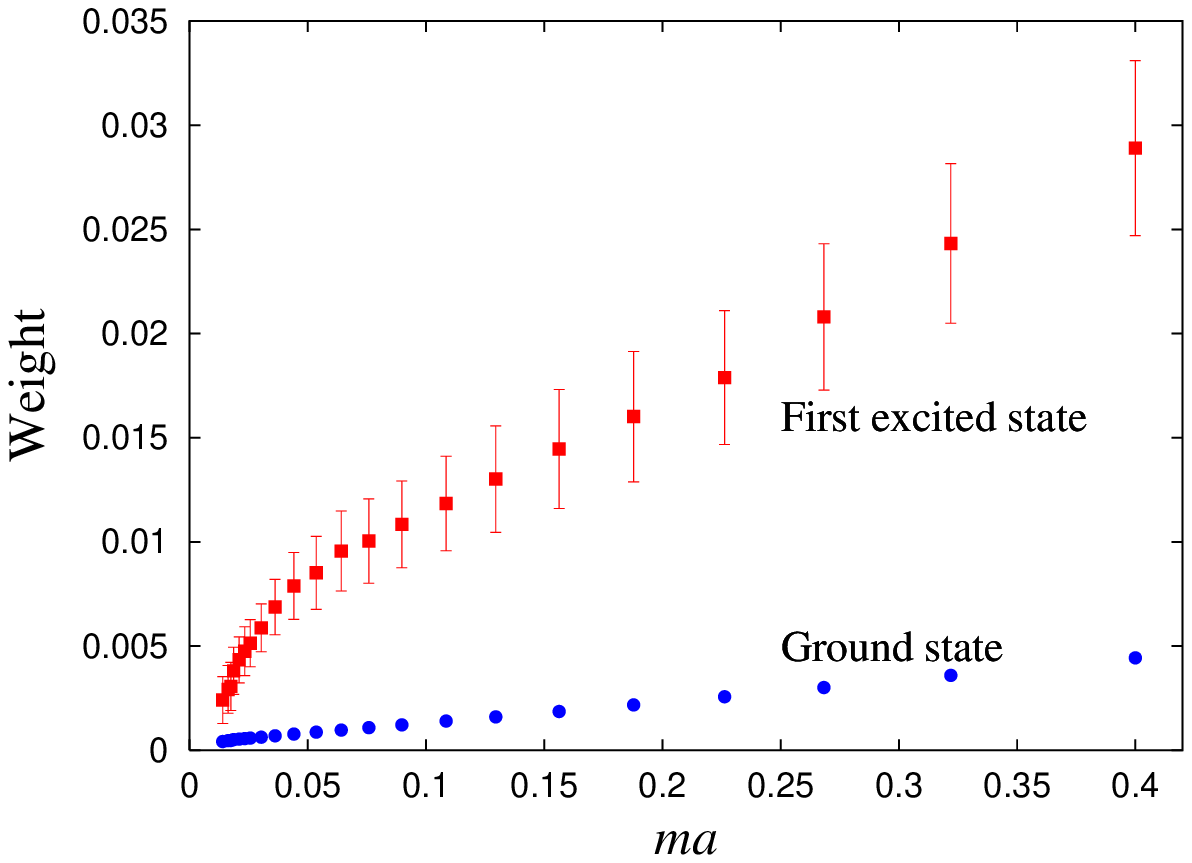}%
\caption{\label{Fig:PionMass} Ground and first-excited state pion mass
$m_{\pi}a$, as a function of the bare quark mass $ma$ (left).  Ground and
first-excited stated pion weight, as a function of the bare quark mass $ma$
(right).  Notice that the ground state weight does not diverge as the quark
mass approaches zero for this $\langle A_4 A_4 \rangle$ correlator as it would
for the $\langle P P \rangle$ correlator, where $P$ is the pseudoscalar
density.}
\end{figure}

In fact, we find that the ``variable $\Delta t$'' method works very well for
the pion correlators.  Figure~\ref{Fig:PionMass} shows the results of the fit,
the ground- and first-excited-state weight and mass as a function of the bare
quark mass.

\subsection{The Roper Resonance} \label{Sect:Roper}

Studies using standard curve fitting have heretofore failed to satisfactorily
identify the ``Roper resonance'' $N^{1/2+}(1440)$ of the nucleon on the
lattice~\cite{Lee:1998cx,Lee:2000hh,Lee:2001ts,Richards:2001bx,Sasaki:2001nf,
Maynard:2002ys,Melnitchouk:2002eg,Sasaki:2002sj, Sasaki:2003xc,Edwards:2003cd}.
Our analysis is the first lattice calculation to obtain the masses of the Roper
and $S_{11}(N^{1/2-}(1535))$ at low quark masses well in the chiral regime
(with a pion mass as low as $180\,{\rm MeV}$).  However, the effects of
quenched artifacts, specifically the presence of ghost $\eta^{\prime}N$ states,
complicates the physics as well as the functional form of fit model, and so the
details of the calculation are presented elsewhere~\cite{Dong:2003zf}.  To be
sure, our calculation benefits from going to unprecedented low quark mass with
the full chiral symmetry provided by overlap fermions, but the Sequential
Empirical Bayes Method plays a crucial role.

\begin{figure}[ht]
\includegraphics[angle=0,width=0.5\hsize]{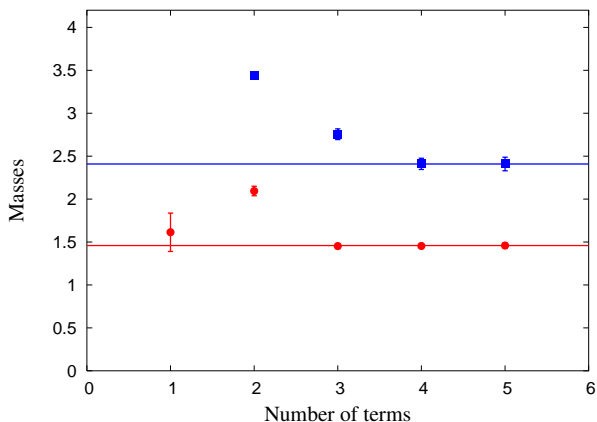}%
\caption{\label{Fig:Roper_stability_previous} Stability test for the Roper
resonance at bare quark mass $ma=0.268$ ($m_{\pi}=702(2)\,{\rm MeV}$).  Priors
had previously been selected by the plain ``variable $\Delta t$'' version of
the Sequential Empirical Bayes Method.  Now they are used in the standard
constrained fit.  The figure shows fit values for the lowest two masses
(Nucleon and Roper) from constrained fits with different numbers of terms in
the fit model. (At this mass, ghost states make no contribution, and so the fit
model contains only exponentials.)  }
\end{figure}

It also presented a challenge and led to the refinement of the SEB from the
``variable $\Delta t$'' version to the ``Just-in-Time'' version.  The first
version of our Roper results~\cite{Dong:2003zfv1} used the ``variable $\Delta
t$'' method which, as we've seen, had until then been perfectly adequate to
expose excited states (such as for the pion in Sec.~\ref{Sect:Pion}) where the
density of excited states is not too high.  Compared to our final analysis with
the ``Just-in-Time'' version~\cite{Dong:2003zf}, the results for the nucleon
and $S_{11}$ did not change.  The only change is for the Roper state at medium
and heavy quark masses.  Indeed our first results at medium-heavy masses passed
the ``Stability Plot'' test (Fig.~\ref{Fig:Roper_stability_previous}) and our
focus returned to the more interesting light quark mass regime.  Here the
``variable $\Delta t$'' criterion for when to add new states remains perfectly
adequate for pion masses below $\sim 350\,{\rm MeV}$.

Subsequently, we came to realize that the ``variable $\Delta t$'' criterion
that we adopted to introduce the excited state was not adequate for medium and
heavy quarks, although it is adequate for the light quarks.  In fitting someone
else's proprietary charmonium data, we realized that the crucial difference
between the heavy quark spectrum and that of the light is that the ratio of the
excited state mass to that of the ground state is much smaller in the heavy
system, that is, the excitation spectrum in the heavy quark system is more
dense.  The earlier criterion erroneously resulted in obtaining the average of
the higher excited states as the first excited state for the heavy system.  In
view of this, we devised the new and final ``Just-in-Time'' criterion
(Sec.~\ref{Sect:FinalAlgorithm}), wherein a new term is added if its
contribution is statistically significant at the current time slice.  In fact,
the penultimate ``Not-too-Soon'' algorithm (Sec.~\ref{Sect:FinalAlgorithm}),
where the criterion for adding a new term to the fit model depends on a sudden
jump in the $\chi^{2}/{\rm dof}$, works just as well as ``Just-in-Time'' for
the Roper resonance at both medium-high and low quark masses.  But for some
charmonium data, or other test data constructed with a dense excited-state
spectrum, ``Just-in-Time'' is somewhat better than ``Not-to-Soon'', and because
of its potential theoretical advantages is our method of choice.

\begin{figure}[ht]
\includegraphics[angle=0,width=0.5\hsize]{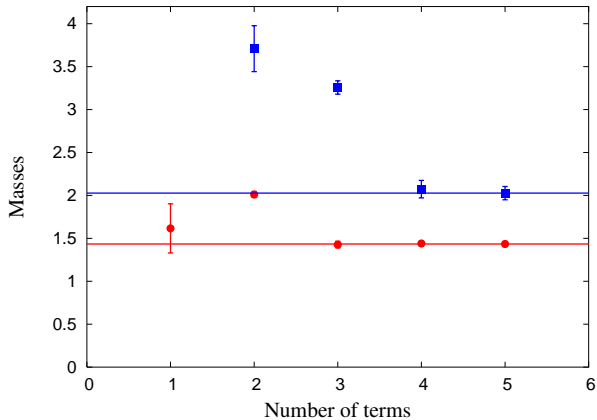}%
\caption{\label{Fig:Roper_stability} Same as for
Fig.~\ref{Fig:Roper_stability_previous} but for our best and final
``Just-in-Time'' algorithm.  }
\end{figure}

Fig.~\ref{Fig:Roper_stability} shows that our final algorithm passes the
``Stability Test''.  However, so did the earlier ``variable $\Delta t$''
algorithm.  This and the fact that the extracted excited state mass changed
alarmingly means that the ``Stability Test''~\cite{Lepage:2001ym} is only a
necessary but not a sufficient test for the reliability of constrained-curve
fitting.  We emphasize, however, that the source of the discrepancy has been
identified and cured; the final algorithm can robustly protect against the
averaging of excited states.

\subsection{Handling Ghost States}

In the quenched approximation, there are artifacts associated with the absence
of quark loops.  One of the more interesting consequences is that the would-be
$\eta^{\prime}$ propagator involves only double $\eta$ poles in ``hairpin
diagrams''.  These lead to the chiral-log terms contributing to hadron masses;
we see these clearly in our recent lattice calculation of pion and nucleon
masses~\cite{Dong:2003im}.  Another quenched artifact is the contribution of
ghost states in the hadron propagators as first seen in the $a_0$ meson
channel~\cite{Bardeen:2001jm} where the ghost $S$-wave $\eta^{\prime}\pi$ state
lies lower in mass than the $a_0$ (for sufficiently small quark mass).  We have
seen a similar effect in analysis of the excited nucleon spectrum where a
$P$-wave $\eta^{\prime}N$ lies close in mass to the Roper
resonance~\cite{Dong:2003zf}, and its presence must be carefully disentangled
by the fitting code.  More dramatically, in the negative-parity channel
$S_{11}$ ($N^{\frac{1}{2}-}$), the lowest $S$-wave $\eta^{\prime}N$ state has a
mass lower than that of the $S_{11}$ for small quark mass.  Since the
$\eta$--$\eta$ coupling in the hairpin diagram is
negative~\cite{Bardeen:2001jm}, the $S_{11}$ correlator changes sign with
increasing time separation~\cite{Dong:2003zf}.  This effect is only seen at
small enough quark masses, and thus is not seen in most lattice simulations at
much higher masses.

As a result, the form of the fitting function changes; there are extra
non-exponential terms.  Nevertheless, the SEB can successfully handle this
situation.  It is crucial to enforce the physical constraint that the weight of
the ghost state be negative; otherwise there would be no way to distinguish it
from the physical state (e.g. Roper) lying nearby.  The details of the fitting
model and the results of the fitting (using the ``Just-in-Time'' SEB algorithm)
for the Roper and $S_{11}$ are in~\cite{Dong:2003zf}.

\begin{figure}[ht]
\includegraphics[angle=0,width=0.45\hsize]{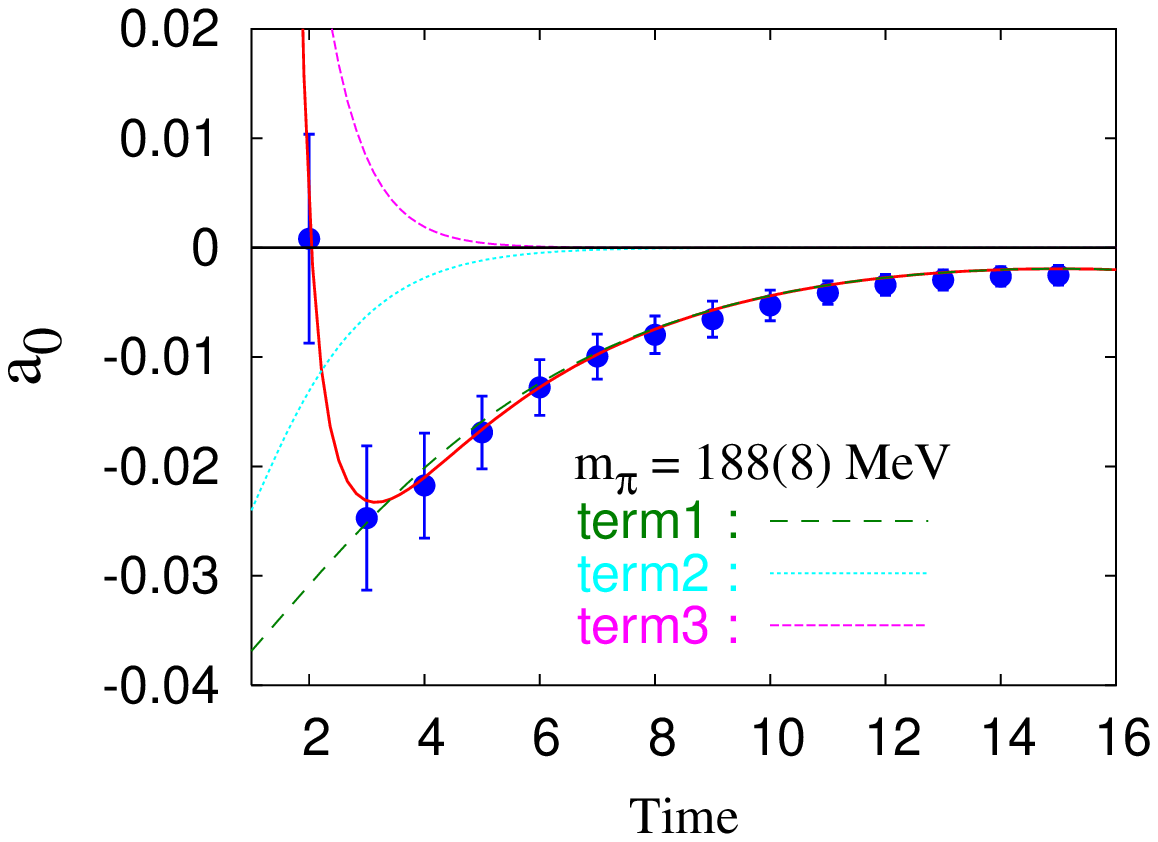}%
\includegraphics[angle=0,width=0.45\hsize]{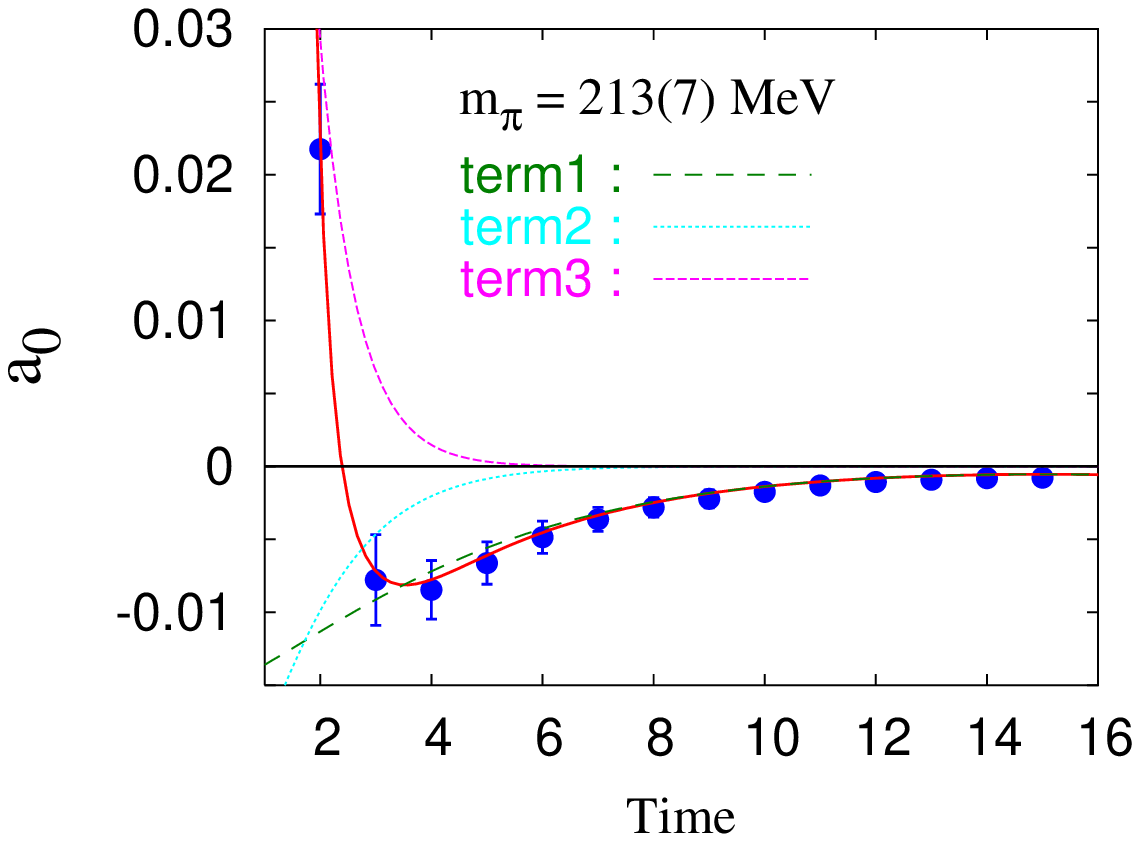}%
\caption{\label{Fig:a0_prop_1} $a_0$ correlators for our lowest quark mass for
which $m_{\pi}=188(8)\,{\rm MeV}$ (left) and $m_{\pi}=213(7)\,{\rm MeV}$
(right).  The negative dip of the correlators is an indication of the
domination of the ghost $S$-wave $\eta^{\prime}\pi$ state over the physical
$a_{0}$.  The curves are contributions to the fit model and are explained in
the text.}
\end{figure}

Another example is displayed in Fig.~\ref{Fig:a0_prop_1} where we show the
$a_0$ propagator at two low quark masses, for which $m_{\pi}=188(8)\,{\rm MeV}$
(left) and $m_{\pi}=213(7)\,{\rm MeV}$ (right).  The most dramatic feature of
the data in each figure is the negative dip of the correlator at time slice 3
and above.  The solid (red) line passing through the data is the result of the
SEB fit.  The two dominant contributions to the fit model are negative and are
indicated by the dashed (green) line labeled ``term 1'' and by the solid (cyan)
curve labeled ``term 2''.  They are modeled by the expression
\begin{equation}
w_1 (1+E_{\pi}t) e^{-m_{\eta^{\prime}\pi} t}
\end{equation}
where $w_1$ is constrained to be negative and the $(1+E_{\pi}t)$ factor
reflects the double-pole nature of the hairpin diagram~\cite{Dong:2003zf}.  We
fit $m_{\eta^{\prime}\pi}$, the mass of the interacting would-be
$\eta^{\prime}$ and $\pi$ state.  Since we work in a finite box, the $\eta'
\pi$ states are discrete and they are constrained to be near the energy of the
two non-interacting particles, each with $E = \sqrt{m^2 +
\sum_i(\frac{2}{a}\sin\left(\frac{p_i}{2})\right)^2}$ for discrete lattice
momentum $p_i = \frac{2\pi n}{L}$ where $n$ is an integer.  For ``term 1'',
$n=0$ and for ``term 2'', $n=1$.  The contribution of the physical state $a_0$
is positive and is indicated by dotted magenta curve in each figure.  In
summary, our SEB algorithm is quite capable of handling non-standard forms of
the fit model including negative-normed ghost-state contributions and can fit
successfully the data.


\section{Summary}

We have advocated refinements of Bayesian-inspired constrained-curve fitting
which we believe better stabilizes fits at low quark mass.  This permits
analysis where the values of fit parameters are not reliably known {\it a
priori}, such that their use as priors might be dangerously biased.

In the ``Sequential Empirical Bayes Method'', we have constructed an automated
and natural way of reliably obtaining the priors from naturally-nested subsets
of the data (``onion shells'').  For the prototype described here, a
time-dependent two-point correlation function $G(t)$, we first obtain estimates
of the ground state mass and weight from unconstrained fits to a subset of data
restricted to large times.  These are then used as the priors in a subsequent
constrained fit.  A sequence of (fully or partially) constrained fits follow.
At each step, more data from adjacent earlier times are added and another term
is added to the fit model.  When a new parameter is added, its first estimate
is obtained from a fit in which it is unconstrained but previously introduced
parameters are constrained.  After each fit, the fitted values of each
parameter are used as the priors for the next fit.  Thus the value of each
prior is free to float from one fit to the next, and thus is not prevented from
changing significantly over the course of the several steps of the iteration as
more and more data are added.  This mitigates any bias which may have been
introduced wherein the first estimates of a prior are inaccurate due to
statistical fluctuations in the smaller initial data set.  In the final step,
as the last data are added, no new terms are added to the fit model.  A final
fit for each parameter leaves it unconstrained (or very weakly constrained)
with all other parameters constrained by priors set equal to the latest fitted
values.  Our final bootstrapped errors obtained from ``releasing the
constraint'' in this way are larger than for a completely-constrained fit and
are selected as a conservative estimate.

We have found some further refinements to be fruitful: (a) We use a
``scanning'' technique to automate the selection of an initial value for a new
parameter for its first unconstrained fit.  We thus avoid the pitfall of the
$\chi^{2}$ minimization routine becoming trapped in the attraction basin of a
local, but not global minimum.  (b) The weight factor (essentially a Lagrange
multiplier), $\lambda$, balances the influence of the data versus the priors in
determining the output of a fit.  Decreasing $\lambda$ from its canonical
Bayesian value of 1, can be used to assess the systematic error associated with
the choice of prior(s).  We advocate absorbing this systematic error into a
statistical error by promoting $\lambda$ to be ``global dynamical weight'',
that is, a fit parameter with its own prior mean and standard deviation.

We have tested that our algorithm can successful recover the correct fit
parameters of an artificial data set, constructed as a sum of decaying
exponentials with realistic values of the parameters.  For one channel
(corresponding to using only the local-local two-point correlation function),
the masses and weights of the ground and first-excited states are fit to within
a measured standard deviation of the true values.  Thus, for our real data, we
restrict ourselves to measuring only the lowest two states, even though we use
four or five states in the fit model.  This is sufficient for our present
purposes.

Our method is not in strict accord with the Bayesian philosophy, as we use
subsets of the data to guide the selection of the priors.  Nevertheless, the
following mitigates bias from such data snooping: (a) The data is naturally
nested in that a subset of data restricted to large times can provide a fair
estimate of the lowest state parameters.  Adiabatically increasing the data set
with the introduction of new terms in the fit model gives fair estimates of
excited-state parameters to be used as priors.  (b) The priors are given ample
opportunity to change in accordance with the data in the several steps of the
algorithm.  Is this enough?  To test this we ``partition the data'' by
configuration into equally-sized but disjoint sets and apply our algorithm to
one half to obtain priors and then use these priors in a standard constrained
fit on the other half (in strict accord with the Bayesian philosophy) and on
the full data set.  We see no difference beyond expected statistical errors;
that is, no bias is seen.

In further studies using artificially constructed data with known means and
errors (``toy models'') we have presented some caveats against the careless
application of the algorithm which might result in a state being spuriously
identified as a combination of two, leading to a ``false positive'' prediction.
These false positives were avoided in the toy model studies by insisting that a
new term be added to the fit model ``Just-in-Time'', that is only after a fit
with fewer terms is deemed inadequate.

For this pilot study, we have analyzed the ground and first-excited masses and
weights for the pion from overlap fermions on a quenched $16^3\times 28$
lattice with spatial size $La=3.2\,{\rm fm}$ and pion mass as low as $180\,{\rm
MeV}$, commented on the history of the use and suitability of variations of the
SEB algorithm for extracting the Roper resonance~\cite{Dong:2003zf}, and given
another example ($a_0$) where the method can handle ghost states.

\begin{acknowledgments}
This work is partially supported by DOE Grants DE-FG05-84ER40154 and
DE-FG02-95ER40907.  We wish to thank A. Alexandru, P. deForcrand, and
L. Glozman.
\end{acknowledgments}


\end{document}